\documentclass[a4paper,UKenglish,cleveref, autoref, thm-restate]{lipics-v2021}
\usepackage{savesym}
\savesymbol{jmath}

\usepackage{sfmath}
\usepackage{breakurl}             
\usepackage{underscore}           

\usepackage[many]{tcolorbox}
\usepackage{mathtools}
\usepackage{amsmath}
\usepackage{bussproofs}
\savesymbol{amalg}
\usepackage[matha]{mathabx}
\usepackage{amssymb, amsthm}
\usepackage{stmaryrd}
\usepackage{float}
\usepackage{amsfonts}
\savesymbol{mathscr}

\usepackage{mathrsfs}
\usepackage[mathscr]{eucal}
\tcbuselibrary{skins}

\definecolor{darkblue}{RGB}{0,60,220}

\newcommand{\WF}{\textup{\texttt{well-formed}}}
\newcommand{\Dedukti}{\textsc{Dedukti}}
\newcommand{\Type}{\textbf{\textup{Type}}}
\newcommand{\Kind}{\textbf{\textup{Kind}}}
\newcommand{\red}{\xhookrightarrow{}}
\newcommand{\invred}{\xhookleftarrow{}}
\newcommand\reds{\mathrel{{\red}{}^*}}
\newcommand\invreds{\mathrel{{}^*{\invred}}}

\newcommand{\trans}[1]{\llbracket #1 \rrbracket}

\newcommand{\lsucc}{\textup{\texttt{s}}}
\newcommand{\lzero}{\textup{\texttt{z}}}

\newcommand{\rewriterules}[1]{}

\newcommand{\UPP}{\textbf{UPP}}
\newcommand{\DkCheck}{\textsc{DkCheck}}
\newcommand{\Agda}{\textsc{Agda}}
\newcommand{\Predicativize}{\textsc{Predicativize}}
\newcommand{\Universo}{\textsc{Universo}}

\newcommand\hide[1]{}
\renewcommand\show[1]{#1}

\newcommand\tolong\show
\newcommand\toshort\hide

\tolong{\hideLIPIcs}

\newcommand\lvleq{\simeq }
\newcommand\teq{\equiv}

\title{Translating proofs from an impredicative type system to a predicative one}

\author{Thiago Felicissimo}{Université Paris-Saclay, INRIA project, Deducteam, Laboratoire Méthodes Formelles, ENS Paris-Saclay, 91190 France}{thiago.felicissimo@inria.fr}{}{}
\author{Frédéric Blanqui}{Université Paris-Saclay, INRIA project, Deducteam, Laboratoire Méthodes Formelles, ENS Paris-Saclay, 91190 France}{thiago.felicissimo@inria.fr}{}{}
\author{Ashish Kumar Barnawal}{Indian Institute of Technology Guwahati, Guwahati 781039, Assam, India}{barnawal@iitg.ac.in}{}{}

\titlerunning{Translating proofs from an impredicative type system to a predicative one}

\nolinenumbers

\toshort{\relatedversion{A long version with all proofs is available at arXiv, at \url{TODO.ADD.LINK}}}
\tolong{\relatedversion{This is the long version of a paper accepted for publication at CSL 2023}}

\authorrunning{T. Felicissimo, F. Blanqui \& A. Kumar Barnawal}
\Copyright{Thiago Felicissimo, Frédéric Blanqui \& Ashish Kumar Barnawal} 

\supplement{\url{https://github.com/Deducteam/predicativize}}

\acknowledgements{The authors would like to thank François Thiré for the help while developing \textsc{Predicativize}, Gilles Dowek for  remarks about previous versions of this paper, Jesper Cockx and Vincent Moreau for discussions about universe levels and the anonymous reviewers for the very helpful comments and remarks.}

\EventEditors{Bartek Klin and Elaine Pimentel}
\EventNoEds{2}
\EventLongTitle{31st EACSL Annual Conference on Computer Science Logic (CSL 2023)}
\EventShortTitle{CSL 2023}
\EventAcronym{CSL}
\EventYear{2023}
\EventDate{February 13--16, 2023}
\EventLocation{Warsaw, Poland}
\EventLogo{}
\SeriesVolume{252}
\ArticleNo{12}

\begin{document}
\maketitle

\begin{abstract}
  As the development of formal proofs is a time-consuming task, it is important to devise ways of sharing the already written proofs to prevent wasting time redoing them. One of the challenges in this domain is to translate proofs written in proof assistants based on impredicative logics, such as \textsc{Coq}, \textsc{Matita} and the \textsc{HOL} family, to proof assistants based on predicative logics like \Agda{}, whenever impredicativity is not used in an essential way.

  In this paper we present an algorithm to do such a translation between a core impredicative type system and a core predicative one allowing prenex universe polymorphism like in \textsc{Agda}. It consists in trying to turn a potentially impredicative term into a universe polymorphic term as general as possible. The use of universe polymorphism is justified by the fact that  mapping an impredicative universe to a fixed predicative one is not sufficient in most cases.

  During the algorithm, we need to solve unification problems modulo the max-successor algebra on universe levels. But, in this algebra, there are solvable problems having no most general solution. We however provide an incomplete algorithm whose solutions, when it succeeds, are most general ones.

  The proposed translation is of course partial, but in practice allows one to translate many proofs that do not use impredicativity in an essential way.  Indeed, it was implemented in the tool \textsc{Predicativize} and then used to translate semi-automatically many non-trivial developments from \textsc{Matita}'s arithmetic library to \Agda{}, including Bertrand's Postulate and Fermat's Little Theorem, which were not available in \Agda{} yet.

\end{abstract}

\section{Introduction}

An important achievement of the research community in logic is the invention of proof assistants. Such tools allow for interactively writing proofs, which are then checked automatically and  can then be reused in other developments. Unfortunately, a proof written in a proof assistant cannot be reused in another one, which makes each tool isolated in its own library of proofs. This is specially the case when considering two proof assistants with incompatible logics, as in this case simply translating from one syntax to another  would not work. Therefore, in order to share proofs between systems it is very often required to do logical transformations.

One approach to share proofs from a proof assistant  $ A $ to a proof assistant $ B $ is to define a transformation acting directly on the syntax of $ A $ and then implement it using the codebase of $ A $. However, this code would be highly dependent on the implementation of $ A $ and can easily become outdated if the codebase of $ A $ evolves. Moreover, if there is another proof assistant $ A' $ whose logic is very similar to the one of $ A $ then this transformation would have to be implemented another time in order to be used with $ A' $.

\vspace{-1em}
\paragraph*{Dedukti} The logical framework \Dedukti{} \cite{dedukti} is a good candidate for a system where multiple logics can be encoded, allowing for logical transformations to be defined uniformly \textit{inside} \Dedukti{}.

Indeed, first, the framework was already shown to be sufficiently expressive to encode the logics of many proof assistants \cite{thU}. Moreover, previous works have shown how proofs can be transformed inside \Dedukti{}. For instance, Thiré describes in \cite{sttfa} a transformation to translate a proof of Fermat's Little Theorem from the Calculus of Inductive Constructions to Higher Order Logic (HOL), which can then be exported to multiple proof assistants such as \textsc{HOL}, \textsc{PVS}, \textsc{Lean}, etc. Géran also used Dedukti to export the formalization of Euclid's Elements Book 1 in \textsc{Coq} \cite{geocoq} to several proof assistants \cite{yoan}.

\vspace{-1em}
\paragraph*{(Im)Predicativity}

One of the challenges in proof interoperability is sharing proofs coming from impredicative proof assistants (the majority of them) to predicative ones such as \Agda{}. Indeed, impredicativity, which is the ability in a logic to quantify over arbitrary entities, regardless of size considerations, is incompatible with predicative systems, in which each entity can only quantify over smaller ones.

Therefore, it is clear that any proof that uses such characteristic in an essential way cannot be translated to a predicative system. Nevertheless, one can wonder if most proofs written in impredicative systems really need impredicativity and, if not, how one could devise a way for detecting  and translating them to predicative systems.

\vspace{-1em}
\paragraph*{Our contribution}

In this paper, we tackle this problem by proposing an algorithm that tries to do precisely this. This algorithm was  implemented on top of the \DkCheck{} type-checker for \Dedukti{} with the tool \textsc{Predicativize}, allowing for the translation of proofs semi-automatically inside \Dedukti{}. These proofs can then be exported to \Agda{}, the main proof assistant based on predicative type theory.

This tool has been used to translate many proofs semi-automatically to \Agda{}, including \textsc{Matita}'s arithmetic library. It contains many-non trivial proofs, and in particular a proof of Bertrand's Postulate which was the  subject of a whole publication \cite{bertrand} -- thanks to our tool, the same hard work did not have to be repeated in order to make it available in \Agda{}. 

\vspace{-1em}
\paragraph*{Outline}

We start in Section \ref{sec:dedukti} with an introduction to \Dedukti, before moving to Section \ref{sec:firstlook}, where we present informally the problems that appear when translating proofs to predicative systems. We then introduce in Section \ref{sec:upp} a predicative universe-polymorphic system, which is a subsystem of \Agda{} and is used as the target of the translation.  This is followed by Section \ref{sec:alg}, the main one, where we present our algorithm. Section \ref{sec:solving} then proposes an (incomplete) unification algorithm for universe levels, which is used by the predicativization algorithm. We then introduce the tool \textsc{Predicativize} in Section \ref{sec:predicativize}, and describe the translation of \textsc{Matita}'s library in Section \ref{sec:matita}. We end with some remarks in Section \ref{sec:conc}. The proofs not given in the main body of the article can be found in the long version (see link on the first page).

\section{Dedukti}
\label{sec:dedukti}

\tolong{
\begin{figure}
{\small\begin{center}
  \AxiomC{}
\RightLabel{\texttt{Empty}}
\UnaryInfC{$-; -~\texttt{well-formed}$}
\DisplayProof
\hskip 1.5em
\AxiomC{$\Sigma;- \vdash A : \textbf{s}$}
\RightLabel{\texttt{Decl-cons}}
\LeftLabel{$c \notin \Sigma$}
\UnaryInfC{$\Sigma, c : A;-~\texttt{well-formed}$}
\DisplayProof
\end{center}
\begin{center}
\AxiomC{$\Sigma;- \vdash M : A$}
\RightLabel{\texttt{Decl-def}}
\LeftLabel{$c \notin \Sigma$}
\UnaryInfC{$\Sigma, c : A := M;-~\texttt{well-formed} $}
\DisplayProof
\hskip 1.5em
\AxiomC{$\Sigma;\Gamma \vdash A : \Type$}
\RightLabel{\texttt{Decl-var}}
\LeftLabel{$x \notin \Gamma$}
\UnaryInfC{$\Sigma;\Gamma, x : A~\texttt{well-formed} $}
\DisplayProof
\end{center}
\begin{center}
\AxiomC{$\Sigma;\Gamma~\texttt{well-formed}$}
\RightLabel{\texttt{Cons}}
\LeftLabel{$c : A\text{ or }c : A := M \in \Sigma$}
\UnaryInfC{$\Sigma;\Gamma \vdash c : A $}
\DisplayProof
\hskip 1.5em
\AxiomC{$\Sigma;\Gamma~\texttt{well-formed}$}
\RightLabel{\texttt{Var}}
\LeftLabel{$ x : A \in \Gamma $}
\UnaryInfC{$\Sigma;\Gamma \vdash x : A  $}
\DisplayProof
\end{center}
\begin{center}
\AxiomC{$ \Sigma;\Gamma~\texttt{well-formed} $}
\RightLabel{\texttt{Sort}}
\UnaryInfC{$\Sigma;\Gamma \vdash \Type : \Kind$}
\DisplayProof
\hskip 1.5em
  \AxiomC{$\Sigma;\Gamma \vdash M : A $}
  \AxiomC{$\Sigma;\Gamma \vdash B : \textbf{s} $}
  \RightLabel{\texttt{Conv}}
  \LeftLabel{$A \equiv B$}
\BinaryInfC{$\Sigma;\Gamma \vdash M : B $}
\DisplayProof
\end{center}
\begin{center}
  \AxiomC{$\Sigma; \Gamma, x : A \vdash B : \textbf{s} $}
\RightLabel{\texttt{Prod}}
\UnaryInfC{$\Sigma;\Gamma \vdash \Pi    x : A . B : \textbf{s}  $}
\DisplayProof
\hskip 1.5em
\AxiomC{$\Sigma;\Gamma \vdash M : \Pi x : A . B  $}
\AxiomC{$\Sigma; \Gamma \vdash N : A $}
\RightLabel{\texttt{App}}
\BinaryInfC{$\Sigma;\Gamma \vdash M N : B\{N/x\} $}
\DisplayProof
\end{center}
\begin{center}
  \AxiomC{$\Sigma; \Gamma, x : A \vdash B : \textbf{s} $}
  \AxiomC{$\Sigma; \Gamma, x : A \vdash M : B $}
  \RightLabel{\texttt{Abs}}
\BinaryInfC{$\Sigma;\Gamma \vdash \lambda x : A . M :\Pi x : A . B  $}
\DisplayProof
\end{center}
\caption{Typing rules for \Dedukti}
\label{typing-dk}}
\end{figure}
}

In this work we use \Dedukti{} \cite{dedukti, thU} as the metatheory in which we express the various logic systems and define our proof transformation. Therefore, we start with a quick introduction to this system. The logical framework \Dedukti{} has the syntax of the $ \lambda $-calculus with dependent types ($ \lambda\Pi $-calculus).
\begin{align*}
  A, B, M, N &::= x \mid c \mid M N \mid\lambda x : A . M\mid\Pi x : A. B \mid \Type \mid\Kind
\end{align*}
Here, $ c $ ranges in a set of constants $ \mathcal{C} $, and $ x $ ranges in an infinite set of variables $ \mathcal{V} $ disjoint from $ \mathcal{C} $.  We call a type of the form $ \Pi x : A.B $ a \textit{dependent product}, and we write $ A \to B $ when $ x $ does not appear free in $ B $. We use  $ \textbf{s} $ to refer to either $ \Type $ or $ \Kind $.

A \textit{context} $ \Gamma $ is a finite sequence of entries of the form $ x : A $. A \textit{signature} $ \Sigma $ is a finite sequence of entries of the form $ c : A $ (constant declarations) or $ c : A := M $ (definitions). It can be useful to split the signature into a global signature $ \Sigma $ and a local signature $ \Delta $ defined on top of the global one. The global signature  holds the definition of the object logic we are working in, whereas the local one holds axioms and definitions inside the logic. For instance, when working with natural numbers in predicate logic we would have $ \land : Prop \to Prop \to Prop \in \Sigma $, as $ \land $ is in the definition of predicate logic, but $ + : Nat \to Nat \to Nat \in \Delta$, given that the natural numbers and addition are not part of predicate logic, but can be defined on top of it.

The main difference between \Dedukti{} and the $ \lambda \Pi $-calculus is that we also consider a set $ \mathscr{R} $ of \textit{rewrite rules}, which are pairs of the form $ c~l_{1}...l_{k} \red r$ where $ l_{1},...,l_{k},r $ are terms. Given a signature $\Sigma,\Delta$, we also consider the $ \delta $ rules allowing for the unfolding of definitions: we have $ c \red M \in \delta$  for each $ c  :A := M \in \Sigma,\Delta $. We then denote by $ \red_\mathscr{R} $ the closure by context and substitution of $ \mathscr{R} $, and by $ \red_\delta $ the closure by context of $ \delta $. Finally, we write $\red_{\beta\mathscr{R}\delta}$ for $\red_\beta  \cup \red_\mathscr{R} \cup \red_\delta$  and $ \equiv_{\beta\mathscr{R}\delta} $ for its reflexive, symmetric and transitive closure.

Rewriting allows us to define equality by computation, but not all equalities  can be defined  like this  in a well-behaved way, e.g. the commutativity of some operator. Therefore, we also consider \textit{rewriting modulo equations} \cite{blanqui03rta}. If $\mathscr{E}$ is a set of pairs of Dedukti terms (written as $M \approx N$), we write $\simeq_{\mathscr{E}}$ for its congruent closure -- that is, its reflexitive, symmetric and transitive closure by context and substitution.

Because $\mathscr{R}$ and $\mathscr{E}$ are usually kept fixed, in the following we write $\red$ for $\red_{\beta\mathscr{R}\delta}$, $\simeq$ for $\simeq_{\mathscr{E}}$ and $\equiv$ for the reflexitive, symmetric and transitive closure of $\red\cup\simeq_{\mathscr{E}}$.

One very important notion that we will use in this work is that of a \textit{theory}, which is a triple $ (\Sigma, \mathscr{R},  \mathscr{E}) $ where $ \Sigma $ is a global signature and  all constants appearing in $ \mathscr{R} $ and $\mathscr{E}$ are declared in $ \Sigma $. Theories are used to define in \Dedukti{} the object logics in which we work (for instance, predicate logic).

The typing rules for  \Dedukti{} are given in \tolong{Figure \ref{typing-dk}}\toshort{Appendix \ref{sec:dktyping}, along with some basic metaproperties that we use in the subsequent proofs}. We remark in particular that the conversion rule of the system allows to exchange types which are equivalent modulo $ \equiv $, which can use not only $\beta$ but also $\delta$, $\mathscr{R}$ and $\mathscr{E}$.

\tolong{
We recall the following basic metaproperties of \Dedukti{}. Proofs can be found in \cite{frederic-phd, saillard15phd}.

\begin{theorem}[Basic metaproperties]
~
  \begin{enumerate}
  \item Weakening:   If $ \Sigma;\Gamma \vdash M : A $, $ \Gamma \subseteq \Gamma' $ and $ \Sigma;\Gamma'~\WF $ then $ \Sigma;\Gamma' \vdash M : A $
  \item Substitution Lemma: If $ \Sigma; \Gamma,x:B,\Gamma' \vdash M : A $ and $ \Sigma;\Gamma \vdash N : B $ then $ \Sigma;\Gamma,\Gamma'\{N/x\}\vdash M\{N/x\} : A\{N/x\} $
  \item Well-sortedness: If $ \Sigma;\Gamma \vdash M : A $ then either $ A = \Kind $ or $ \Sigma;\Gamma \vdash A : \textup{\textbf{s}} $ for $ \textbf{\textup{s}}  = \Type$ or $ \Kind $.
  \item Subject reduction of $ \delta $: If $ \Sigma;\Gamma \vdash M : A $ and $ M \red_\delta M' $ then $ \Sigma;\Gamma \vdash M' : A $
  \item Subject reduction of $ \beta $: If injectivity of dependent products holds, then $ \Sigma;\Gamma \vdash M : A $ and $ M \red_\beta M' $ implies $ \Sigma;\Gamma \vdash M' : A $.
  \item Contexts are well typed: If $ x : A \in \Gamma $ then $ \Sigma;\Gamma \vdash A : \Type $
  \item Signatures are well typed: If $ c : A \in \Sigma$  then $ \Sigma;- \vdash A : \textup{\textbf{s}}$ and if $ c : A := M \in \Sigma $ then $  \Sigma;-\vdash M : A$
  \item Inversion of typing: Suppose $ \Sigma;\Gamma \vdash M : A $
    \begin{itemize}
    \item If $ M = x $ then $ x : A' \in \Gamma $ and $ A \equiv A' $
    \item If $ M = c $ then $ c : A'  \in \Sigma $ and $ A \equiv A' $
    \item If $ M = \Type $ then $ A \equiv \Kind $
    \item  $ M= \Kind $ is impossible
    \item If $ M = \Pi x : A_1. A_2 $ then $ \Sigma; \Gamma,x:A_1 \vdash A_2 : \textbf{\textup{s}} $ and $ \textbf{\textup{s}} \equiv A $
    \item If $ M = M_1 M_2 $ then $ \Sigma; \Gamma \vdash M_1 : \Pi x: A_1.A_2 $, $ \Sigma;\Gamma \vdash M_2 : A_1 $ and $ A_2\{M_2/x\} \equiv A $
    \item If $ M = \lambda x : B. N $ then  $ \Sigma; \Gamma, x:B \vdash C:\textbf{\textup{s}} $, $ \Sigma;\Gamma,x:B \vdash N:C $ and $ A \equiv \Pi x:B.C $
    \end{itemize}
    \item Weak permutation: If $\Sigma; \Gamma, x :A, y : B, \Gamma' \vdash M : C$ and $\Gamma \vdash B : \Type $ then $\Sigma; \Gamma,  y : B, x :A,\Gamma' \vdash M : C$.
  \end{enumerate}
\end{theorem}
}

\subsection{Defining Pure Type Systems in Dedukti}
\label{subsec:pts}

We briefly review how Pure Type Systems (PTSs) \cite{pts} can be defined in \Dedukti{} \cite{dowek2007} (other approaches also exists, such as \cite{felicissimo:LIPIcs.FSCD.2022.25}), as we will need this in the rest of the article. Recall that in PTSs,  universes and function types can be specified by a set $ \mathcal{S} $ of universes, and two relations  $ \mathcal{A}\subseteq\mathcal{S}^2$ and $ \mathcal{R}\subseteq\mathcal{S}^3$ -- which we suppose to be functional relations here, as is usually the case. These specify that, if $ (s_1,s_2) \in \mathcal{A} $, then $ s_1 $ is of type $ s_2 $, and if  $ (s_1,s_2,s_3) \in \mathcal{R} $ then when $ A : s_1$ and $  B :s_2 $ we have $  \Pi x : A. B : s_3 $. Given a PTS specification $ (\mathcal{S},\mathcal{A},\mathcal{R}) $, we can define the corresponding PTS with a \Dedukti{} theory in the following manner.

We first start with the definition of universes. For each universe $ s \in \mathcal{S} $, we declare a \Dedukti{} type $ U_s : \Type$  holding the types in the universe $ s $. We then also declare a function symbol $ El_s : U_s \to \Type $ mapping each member of $ U_s $ to the type of its elements. We might see the elements of $ U_s $ as the codes for the types in $ s $, and $ El_s $ as the decoding function, mapping a code to its true type.

In order to represent the fact that a universe $ s_1 $ is a member of $ s_2 $ when $ (s_1,s_2) \in \mathcal{A} $ we add the constant $ u_{s_1} : U_{s_2} $. However, now the universe $ s_1 $ is represented both by $El_{s_2}~u_{s_1}$ and $U_{s_1}$. Therefore, we add the rewrite rule $ El_{s_2}~u_{s_1} \red U_{s_1} $, stating that $ u_{s_1} $ decodes to $ U_{s_1} $.

Finally, to define dependent functions, for each $ (s_1,s_2,s_3) \in \mathcal{R} $ we add a symbol $ \pi_{s_1,s_2} : \Pi A:U_{s_1}. (El_{s_1}~A \to U_{s_2}) \to U_{s_3} $. Intuitively, the type $ El_{s_3}~(\pi_{s_1,s_2}~A~(\lambda x.B)) $ should hold the functions from $ x:El_{s_1}~A $ to $ El_{s_2}~B $, where $ x $ might occur in $ B $. To make this representation explicit, we add a rewrite rule $ \pi_{s_1,s_2}~A~B \red \Pi x : El_{s_1}~A. El_{s_2}~(B~x) $. Because the type of functions from $ x:A $ to $ B $ is now represented by the framework's function type, the framework's abstraction and application can be used to represent the ones of the encoded system.

In the following, we allow ourselves to write $ \pi_{s_1,s_2}~A~(\lambda x. B) $ informally as $ \pi_{s_1,s_2}~x : A. B $ in order to improve clarity. When $ x \not \in FV(B) $, we might also write $ A \leadsto_{s_1,s_2} B $.

\begin{figure}
\noindent\parbox{.5\textwidth}{
  \begin{align*}
  &U_s :  \Type &\text{for }s \in \mathcal{S}\\
  &El_s :U_s \to \Type &\text{for }s \in \mathcal{S}
\end{align*}}
\parbox{.5\textwidth}{
  \begin{align*}
    &u_{s_1} : U_{s_2}&\text{for } (s_1,s_2) \in \mathcal{A}\\
  &El_{s_2}~u_{s_1} \red U_{s_1}&\text{for } (s_1,s_2) \in \mathcal{A}
  \end{align*}}
\vspace{-2.3em}
\begin{align*}
  &\pi_{s_1,s_2} : \Pi A : U_{s_1}. (El_{s_1}~A \to U_{s_2}) \to U_{s_3} &\text{for }(s_1,s_2,s_3) \in \mathcal{R}\\
  &El_{s_3}~(\pi_{s_1,s_2}~A~B) \red \Pi x : El_{s_1}~A. El_{s_2}~(B~x)&\text{for }(s_1,s_2,s_3) \in \mathcal{R}
\end{align*}\vspace{-2em}                                                                         
\caption{The \Dedukti{} theory which defines the PTS specified by $ (\mathcal{S} , \mathcal{A}, \mathcal{R})$}
\end{figure}

\section{An informal look at the challenges of proof predicativization}
\label{sec:firstlook}

In this informal section we present the problem of proof predicativization and discuss the challenges that arise through the use of examples. Even though the examples might be unrealistic, they showcase real problems we found during our first predicativization attempt, of Fermat's little theorem library in HOL \cite{sttfa} -- some of them being already noted in \cite{delort:hal-02985530}.

We first start by defining the theories \textbf{I} and \textbf{P}, which we will use to represent the core logics of impredicative and predicative proof assistants. These theories are defined as Pure Type Systems as explained in Subsection \ref{subsec:pts} and are described by the specifications bellow. Remember that a universe $ s $ is said  to be impredicative when it is closed under dependent products indexed by some bigger sort, that is, for some $ s' $ with $ (s, s') \in \mathcal{A} $  we have $ (s', s, s) \in \mathcal{R} $. Therefore,  $ \textbf{I} $ is an impredicative system and $ \textbf{P} $ is a predicative one.

\noindent\parbox{.5\textwidth}{
\begin{align*}
  \mathcal{S}_{\boldsymbol{I}} &= \{*, \square\}\\
  \mathcal{A}_{\boldsymbol{I}} &= \{(*, \square)\}  \\
  \mathcal{R}_{\boldsymbol{I}} &= \{(*, *, *), (\square, *, *), (\square,\square,\square)\}
\end{align*}
}
\parbox{.5\textwidth}{
\begin{align*}
  \mathcal{S}_{\boldsymbol{P}} &= \mathbb{N} \\
  \mathcal{A}_{\boldsymbol{P}} &= \{(n, n+1) \mid n \in \mathbb{N} \}  \\
  \mathcal{R}_{\boldsymbol{P}} &= \{(n, m, max\{n,m\}) \mid n,m \in \mathbb{N}\}
\end{align*}  
}

In this setting, the problem of proof predicativization consists in defining a transformation such that, given a local signature $ \Delta $ with $ \Sigma_I, \Delta;-~\texttt{well-formed} $, allows to translate it to a local signature $ \Delta' $ with $ \Sigma_P,\Delta';-~\texttt{well-formed} $. Stated informally, we would like to translate constants (which represent axioms) and definitions (which also represent proofs) from \textbf{I} into \textbf{P}. Note in particular that such a transformation is not applied to a single term but to a sequence of constants and definitions, which can be related by dependency -- this dependency turns out to be a major issue as we will see. In the following we represent the local signature $ \Delta $ in a more readable way as a list of entries $\textbf{constant}~c : A$ and $ \textbf{definition}~c : A := M $.

Now that our basic notions are explained, let us dive into proof predicativization. For our first step,  consider a very simple development showing that for every type $ P $ in $ * $ we can build an element of $ P \leadsto_{*,*} P $ -- if  $ * $ is a universe of propositions, then this is just a proof that each proposition in $ * $ implies itself.
\begin{flalign*}
  &\textbf{definition}~thm_1 : El_{*}~(\pi_{\square,*}~P :u_{*}.P \leadsto_{*,*}P)  := \lambda P : U_{*}. \lambda p : El_*~P. p&
\end{flalign*}%

To translate this simple development, the first idea that comes to mind is to define a mapping on universes: the universe $ * $ is mapped to $ 0 $ and the universe $ \square $ is mapped to $ 1 $. However, because our syntax in Dedukti is heavily annotated, we should only apply this map to the constants $u$ and $U$, which represent the universes, and then try to recalculate the annotations of the other constants $El$ and $ \pi$ (remember that $\leadsto$ is just an alias for $\pi$). This would then yield the following local signature, which is indeed valid in \textbf{P}.
\begin{flalign*}
  &\textbf{definition}~thm_1 : El_{1}~(\pi_{1,0}~P :u_{0}.P \leadsto_{0,0} P) := \lambda P : U_{0}. \lambda p : El_0~P. p&
  \end{flalign*}%

This naive approach however quickly fails when considering other cases. For instance, suppose now that one adds the following definition -- once again, if $ *$ is a universe of propositions, then this is just a proof of  the proposition $ (\forall P . P \Rightarrow P) \Rightarrow \forall P . P \Rightarrow P $.
\begin{flalign*}
  &\textbf{definition}~thm_3 : El_*~((\pi_{\square,*}~P:u_*. P \leadsto_{*,*}P) \leadsto_{*,*} \pi_{\square,*}~P:u_*. P \leadsto_{*,*}P) \\ &\hspace{5em} := thm_1~(\pi_{\square,*}~P:u_*. P \leadsto_{*,*}P)&
\end{flalign*}%

If we try to perform the same syntactic translation as before, we get the following result:
\begin{flalign*}
  &\textbf{definition}~thm_3 : El_1~((\pi_{1,0}~P:u_0. P \leadsto_{0,0}P) \leadsto_{1,1} \pi_{1,0}~P:u_0. P \leadsto_{0,0}P) \\&\hspace{5em}:= thm_1~(\pi_{1,0}~P:u_0. P \leadsto_{0,0}P)&
\end{flalign*}%

However, one can verify that this term is not well typed. Indeed, in the original term one quantifies over all types in $ * $ in the term $ \pi_{\square,*}~P:u_*. P \leadsto_{*,*}P $, and because of impredicativity this term stays at $ * $. However, in \textbf{P} quantifying over all elements of the universe $ 0 $ in $ \pi_{1,0}~P:u_0. P \leadsto_{0,0}P $ raises the type to the universe $ 1 $. As $ thm_1 $ expects a term in the universe $ 0 $, the term $ thm_1~(\pi_{1,0}~P:u_0. P \leadsto_{0,0}P) $ is not well-typed.

This suggests that impredicativity introduces a kind of \textit{typical ambiguity}, as it allows us to hide in a single universe $ * $ all kinds of bigger types which would have to be placed in bigger universes in a predicative setting. Hence, in order to handle cases like this one, which arise a lot in practice, we should not translate every occurrence of $ * $ as $ 0 $ naively as we did, but try to compute for each occurrence of $ * $ some natural number $ i $ such that replacing it by $ i $ would produce a valid result.

Thankfully, performing such kind of transformations is exactly the goal of \textsc{Universo} \cite{thire}. This tool allows one to transport typing derivations between two PTS specifications.

To understand how this works, let us come back to the previous example. \Universo{} starts here by replacing all sorts by occurrences of $ l $, where $ l $ is a fresh metavariable representing a natural number.
\begin{flalign*}
  &\textbf{definition}~thm_1 : El_{l_1}~(\pi_{l_2,l_3}~P:u_{l_4}.P \leadsto_{l_5,l_6} P) := \lambda P : U_{l_7}. \lambda p : El_{l_8}~P. p&\\
  &\textbf{definition}~thm_3 : El_{l_{9}}~((\pi_{{l_{10}},{l_{11}}}~P:u_{l_{12}}. P \leadsto_{{l_{13}},{l_{14}}} P) \leadsto_{{l_{15}},{l_{16}}} \pi_{{l_{17}},{l_{18}}}~P:~u_{l_{19}}.P \leadsto_{{l_{20}},{l_{21}}} P) \\ &\hspace{5em} := thm_1~(\pi_{{l_{22}},{l_{23}}}~P:u_{l_{24}}. P \leadsto_{{l_{25}},{l_{26}}} P)&
\end{flalign*}%

These of course are not valid proofs in \textbf{P}, but in the following step \Universo{} typechecks such development and generates constraints in the process. These constraints are then given to a SMT solver, which is used to compute for each metavariable $ l $ a natural number so that the local signature is valid in \textbf{P}. For instance, applying \Universo{} to our previous example would produce the following valid local signature in \textbf{P}.
{\begin{flalign*}
  &\textbf{definition}~thm_1 : El_{2}~(\pi_{2,1}~P:u_{1}. P \leadsto_{1,1} P)) := \lambda P : U_{1}. \lambda p : El_{1}~P. p&\\
&\textbf{definition}~thm_3 : El_{1}~((\pi_{1,0}~P:u_{0}.P \leadsto_{0,0} P) \leadsto_{1,1} \pi_{1,0}~P:u_{0}. P \leadsto_{0,0} P)  \\&\hspace{5em}:= thm_1~(\pi_{1,0}~P:u_{0}. P \leadsto_{0,0} P)&
\end{flalign*}}%

By using \textsc{Universo} it is possible to go much further than with the naive syntactical translation. Still, this approach also fails when being employed with real libraries. To see the reason, consider the following minimum example, in which one uses an element of $ \pi_{\square,*}~P:u_*.P \leadsto_{*,*} P $ twice to build another element of the same type.
{\begin{flalign*}
  &\textbf{definition}~thm_1 : El_*~(\pi_{\square,*}~P:u_*. P \leadsto_{*,*} P) := \lambda P : U_*. \lambda p : El_*~P. p&\\
  &\textbf{definition}~thm_2 : El_*~(\pi_{\square,*}~P:u_*. P \leadsto_{*,*} P) := thm_1~(\pi_{\square,*}~P:u_*. P \leadsto_{*,*} P)~thm_1&
\end{flalign*}}%

If we repeat the same procedure as before, we get the following term, which generates unsolvable constraints.
{\begin{flalign*}
  &\textbf{definition}~thm_1 : El_{l_{1}}~(\pi_{l_{2},l_{3}}~P:u_{l_{4}}. P \leadsto_{l_{5},l_{6}} P) := \lambda P : U_{l_{7}}. \lambda p : El_{l_{8}}~P. p&\\
  &\textbf{definition}~thm_2 : El_{l_{9}}~(\pi_{l_{10},l_{11}}~P:u_{l_{12}}.P \leadsto_{l_{13},l_{14}} P) := thm_1~(\pi_{l_{15},l_{16}}~P:u_{l_{17}}. P \leadsto_{l_{18},l_{19}} P)~thm_1&
\end{flalign*}}%

This happens because the application $ thm_1~(\pi_{l_{15},l_{16}}~P:u_{l_{17}}. P \leadsto_{l_{18},l_{19}} P)~thm_1 $  forces $ l_4 $ to be both $ l_{17} $ and $ l_{17} +1  $, which is impossible. This example suggests that impredicativity does not only hide the fact that types are stratified, but also the fact that they can be used at any level of this stratification. For instance, in our example we would like to use $ thm_1 $ one time with  $ l_4 = l_{17} $ and another time with $ l_4 = l_{17+1} $. In general, when trying to translate libraries using \textsc{Universo} we found that at very early stages a translated proof or object was already needed at multiple universes at the same time, causing the translation to fail.

Therefore, in order to properly compensate for the lack of impredicativity, our solution uses \textit{universe polymorphism}, a feature in type theory (and also present in \textsc{Agda}) that allows defining terms that can later be used at multiple universes \cite{typechecking-with-universes,coq}. Our translation works by trying to compute for each definition or declaration its most general universe polymorphic type, and using it when translating the subsequent declarations or definitions. To understand how this is done precisely, let us first introduce universe polymorphism, which is the subject of the following section.

\section{A Universe-Polymorphic Predicative Type System}
\label{sec:upp}

In this section, we define the Universe-Polymorphic Predicative Type System (or just \UPP{}), which enriches the Predicative PTS \textbf{P} with prenex universe polymorphism \cite{typechecking-with-universes,guillaume}. This is in particular a subsystem of the one underlying the \textsc{Agda} proof assistant \cite{agda}. As usual, we define this system as a \Dedukti{} theory $ \textbf{UPP}=(\Sigma_{UPP},\mathscr{R}_{UPP},\mathscr{E}_{UPP}) $.

The main change with respect to \textbf{P} is that, instead of indexing the constants $ El_s, U_s, u_s, \pi_{s_1,s_2} $ externally, we index them inside  the framework \cite{assaf}. To do this, we first introduce a syntax for \textit{universe levels} inside \Dedukti{} by the following grammar \[
l, l' ::= i \in \mathcal{I} \mid \lzero \mid \lsucc~l \mid l \sqcup l'
\]where the constants $\lzero, \lsucc$ and $\sqcup$ are defined bellow and $\mathcal{I}\subsetneq \mathcal{V}$ is a set of level variables. We also enforce that level variables can only be substituted by other levels.

\noindent\parbox{.36\textwidth}{
  \begin{align*}
  &Level : \Type\\
  &\lzero : Level
\end{align*}}
\parbox{.5\textwidth}{
  \begin{align*}
  &\lsucc : Level \to Level\\
  &\sqcup : Level \to Level \to Level\quad\text{(written infix)}
  \end{align*}}

The definitions in the theory \textbf{P} of $ El_s, U_s, u_s, \pi_{s_1,s_2} $ and the related rewrite rules are then replaced by the following ones.\footnote{Note that in the following rewrite rules we do not need to impose  $i'$ to be equal or convertible to $\lsucc~i$ or $i_{A}\sqcup i_{B}$, given that, for well-typed instances of the rule, this is ensured by typing \cite{blanqui05mscs, assaf, saillard2015type, blanqui2020type}.}

\noindent\parbox{.36\textwidth}{
  \begin{align*}
  &U : Level \to \Type\\
  &El : \Pi i : Level. U~i \to \Type\\
  &u : \Pi i : Level. U~(\lsucc~i)
\end{align*}}
\parbox{.5\textwidth}{
  \begin{align*}
    &\pi : \Pi (i_A~i_B : Level)~(A : U~i_A). (El~i_A~A \to U~i_B) \to U~(i_A \sqcup i_B)\\
  &El~i'~(u~i) \red U~i\\    
  &El~i'~(\pi~i_A~i_B~A~B) \red \Pi x : El~i_A~A. El~i_B~(B~x)
  \end{align*}}

We however  still allow ourselves to write $ El_l, U_l, u_l,\pi_{l,l'} $ in order to improve clarity. We also reuse the previous convention to write $ \pi_{l,l'}~A~(\lambda x.B) $ as $ \pi_{l,l'}~x:A.B $, or even $ A \leadsto_{l,l'} B $ when $ x \not \in FV(B) $.

Now, (prenex) universe polymorphism can be represented directly with the use of the framework's function type \cite{assaf}. Indeed, if a definition contains free level variables, it can be made universe polymorphic by abstracting over such variables. The following example illustrates  this.

\begin{example}
The universe polymorphic identity function  is given  by  \[
id = \lambda (i : Level). \lambda (A : U_i). \lambda (a : El_i~A). a 
\] which has type $ \Pi i : Level. El_{(\lsucc~i)}~(\pi_{ (\lsucc~i),i}~A : u_i. A \leadsto_{i,i}A)$. This then allows to use $ id $ at any universe level: for instance, we can obtain the polymorphic identity function at the level $ \lzero$ with the application $ id~\lzero$, which has type $ El_{(\lsucc~\lzero)}~(\pi_{(\lsucc~\lzero),\lzero}~A :u_\lzero. A \leadsto_{\lzero,\lzero}A) $.
\end{example}

Finally, in order to finish our definition we need to specify the definitional equality satisfied by levels, which is the one generated by the following equations \cite{agda}. Note that, as stated before, we enforce that $i, i_{1},i_{2},i_{3} \in \mathcal{I}$ can only be replaced by other levels.
\begin{align*}
  &i_{1}\sqcup (i_{2} \sqcup i_{3}) \approx (i_{1}\sqcup i_{2}) \sqcup i_{3} &&\lsucc~(i_{1} \sqcup i_{2}) \approx \lsucc~i_{1} \sqcup \lsucc~i_{2} &&i \sqcup \lzero \approx i\\
  &i_{1}\sqcup i_{2}\approx i_{2} \sqcup i_{1} &&i \sqcup \lsucc~i \approx \lsucc~i &&i \sqcup i \approx i
\end{align*}

This definition is justified by the following property. Given a function $\sigma : \mathcal{I} \to \mathbb{N}$, define the interpretation $\trans{l}_{\sigma}$ of a level $ l $ by interpreting the symbols $ \lzero, \lsucc $ and $ \sqcup $  as zero, successor and max, and by interpreting each variable $i$ by $\sigma(i)$.

\begin{proposition}\label{semanticlvl}
We have $l_{1} \simeq l_{2}$ iff $\trans{l_{1}}_{\sigma} = \trans{l_{2}}_{\sigma}$ holds for all $\sigma$.
\end{proposition}
\tolong{
\begin{proof}
  Note that for each $l \approx l' \in \mathcal{E}_{UPP}$ we have $\trans{l}_{\sigma}=\trans{l'}_{\sigma}$ for all $\sigma$, and thus the direction $\Rightarrow$ can be showed by an easy induction on $l_{1}\simeq l_{2}$.

  For the other direction, suppose that we have $\trans{l_{1}}_{\sigma} = \trans{l_{2}}_{\sigma}$ for all $\sigma$, and let us show $l_{1}\simeq l_{2}$. First note that we can show $l \sqcup \lsucc^{n}~l \simeq \lsucc^{n}~l$ for all $n, l$ by induction on $n$. Using this identity and the others in $\mathcal{E}_{UPP}$, we can show that any level $l$ is related by $\simeq$ to a level $\hat{l}$ of the form $\lsucc^{k}~\lzero\sqcup\lsucc^{n_{i_{1}}}~i_{1}\sqcup ... \sqcup \lsucc^{n_{i_{p}}}~i_{p}$, where $i_{1}...i_{p}$ are different variables and $n_{i_{m}} \leq k$ for all $m = 1,...,p$. By doing this for $l_{1}$ and $l_{2}$, we get $l_{1} \simeq \hat{l_{1}}$ and $l_{2} \simeq \hat{l_{2}}$, and thus $\trans{\hat{l_{1}}}_{\sigma}=\trans{\hat{l_{2}}}_{\sigma}$ for all $\sigma$. By varying $\sigma$ over suitable functions we can show that their normal forms are equal up to reordering, and thus are identified by $\simeq$.
\end{proof}}

This also shows that our definition of $\simeq$ agrees with the one used in other works about universe levels \cite{guillaume, gaspard, blanqui22fscd}. The following basic properties show that $\red$ and $\simeq$ interact well.

\begin{proposition}~
  \begin{enumerate}
  \item $\red$ is confluent
  \item If $M \simeq N \red N'$ then, for some $M'$, we have $M \red M' \simeq N'$.
  \item If $M \equiv N$ then $M \reds M' \simeq N' \invreds N$.
\end{enumerate}
\end{proposition}

\tolong{
\begin{proof}
\begin{enumerate}
  \item  Follows from the fact that our rewrite rules define an orthogonal combinatory rewrite system \cite{CRS}.
  \item By induction on the rewrite context of $N \red N'$. The induction steps are easy, we only show the base cases.
        \begin{enumerate}
          \item If $N = El~j~(u~i) \red U~i$, then we have $M = El~j'~(u~i')$ with $j \simeq j'$ and $i \simeq i'$. Therefore, $El~j'~(u~i')\red U~i' \simeq U~i$.
          \item If $N = El~j~(\pi~i_{1}~i_{2}~A~B) \red \Pi x : El~i_{1}~A.El~i_{2}~(B~x)$, then we have $M = El~j'~(\pi~i'_{1}~i'_{2}~A'~B')$, with $P \simeq P'$ for $P=j,i_{1},i_{2},A,B$. Therefore $El~j'~(\pi~i'_{1}~i'_{2}~A'~B') \red \Pi x : El~i'_{1}~A'.El~i'_{2}~(B'~x) \simeq \Pi x : El~i_{1}~A.El~i_{2}~(B~x)$.
          \item If $N$ is a $\beta$-redex, we have two possibilities.

                \begin{enumerate}
                  \item $N = (\lambda x . P_{1}) P_{2} \red P_{1}\{P_{2}/x\}$ with $x \not \in \mathcal{I}$. Then $M = (\lambda x. P_{1}')P_{2}'$ with $P_{1} \simeq P_{1}'$ and $P_{2}\simeq P_{2}'$, and we can show $P_{1}\{P_{2}/x\} \simeq P'_{1}\{P'_{2}/x\}$. Therefore $(\lambda x. P_{1}')P_{2}' \red P_{1}'\{P_{2}'/x\} \simeq P_{1}\{P_{2}/x\}$.
                  \item $N = (\lambda i. P) l \red P\{l/i\}$ where $i \in \mathcal{I}$. Therefore $M = (\lambda i. P')l'$ with $P \simeq P'$ and $l \simeq l'$. Moreover, because $i$ is a level variable, and we suppose that only levels can be replaced for level variables, $l$ must be a level, and so $l'$. Using this, we can show $P\{l/i\}\simeq P'\{l'/i\}$. Therefore, $(\lambda i. P')l' \red P'\{l'/i\} \simeq P\{l/i\}$.
                \end{enumerate}
        \end{enumerate}

  \item If $M \equiv N$, then we have $M(\simeq(\red \cup \invred)^{*})^{n}N$ for some $n$. We prove the result by induction on $n$, the base case being trivial. For the inductive step, we have \[M (\simeq(\red \cup \invred)^{*})^{n}P(\simeq(\red \cup \invred)^{*}) N\]  for some $P$. By confluence of $\red$ we have  $P \simeq ~ \reds \invreds N$ and thus by iterating $(2)$ we get $P \reds ~\simeq ~\invreds N$. By IH we have $M \reds ~ \simeq ~ \invreds P $. Then, joining the rewrite sequences gives $M \reds~\simeq~\invreds\reds~\simeq~\invreds N$. By confluence another time we have $M \reds~\simeq~\reds\invreds~\simeq~\invreds N$. Now it suffices to iterate $(2)$ once again to conclude. \qedhere

\end{enumerate}
\end{proof}
}

Using the third property, one can apply known techniques to show that $\red$ satisfies subject reduction \cite{blanqui2020type, saillard15phd} (this can also be automatically verified using \textsc{DkCheck} or \textsc{Lambdapi}).

\begin{proposition}
If $\Sigma_{UPP}, \Delta; \Gamma \vdash M : A$ and $M \red M'$ then $\Sigma_{UPP}, \Delta;\Gamma \vdash M' : A$.
\end{proposition}

The third property is also very important from a practical perspective: it shows that in order to check $M \equiv N$ one does not need to use matching modulo $\simeq$.

\section{The algorithm}
\label{sec:alg}

We are now ready to define the (partial) translation of a local signature $ \Delta $ to the theory $ \textbf{UPP} $. The idea of the translation is that we traverse the signature $ \Delta $ and at each step we try to compute the most general universe polymorphic version of a definition or constant. The result of a previously translated definition or declaration can then be used at multiple levels for translating entries occurring later in the signature. In order to understand all the following steps intuitively, we will make use of a running example.

\begin{example}
  The last example of Section \ref{sec:firstlook} corresponds to the local signature
{\begin{align*}
         \Delta_I=&~thm_1 : El_*~(\pi_{\square,*}~P:u_*. P \leadsto_{*,*} P) := \lambda P : U_*. \lambda p : El_*~P. p~,&\\
                  &~thm_2 : El_*~(\pi_{\square,*}~P:u_*. P \leadsto_{*,*} P) := thm_1~(\pi_{\square,*}~P:u_*. P \leadsto_{*,*} P)~thm_1&
\end{align*}}%
which is well-formed in the theory $ \textbf{I} $. Let us suppose that the first entry of the signature has already been translated, giving the following signature $ \Delta_{thm_1} $.
{\begin{align*}
    \Delta_{thm_1}=&~thm_1 : \Pi i : Level.El_{(\lsucc~i)}~(\pi_{(\lsucc~i),i}~P:u_i. P \leadsto_{i,i} P) := \lambda i:Level.\lambda P : U_i. \lambda p : El_i~P. p
       \end{align*}}%
Therefore, as a running example, we will translate step by step the second entry $ thm_2 $.
\end{example}

Let us start with some basic auxiliary definitions. Given a local signature $ \Delta $ such that $ \Sigma_{UPP},\Delta;-~\texttt{well-formed} $ and a constant $ c $ occurring in $ \Delta $, let us define $ \textsc{Arity}(c) $ as the greatest natural number $ k $ such that the type of $ c $  is of the form $ \Pi i_1~..~i_k:Level. A $.  Informally, it is the number of level arguments that this constant expects. For instance, we have  $ \textsc{Arity}(thm_1)=1 $.

Using this function, let us define $ \textsc{InsertMetas}(M) $, by the following equations. This function allows us to insert the fresh level variables that will be used to compute the constraints. We suppose that the inserted variables come from a dedicated subset of level variables $ \mathcal{M} \subsetneq \mathcal{I} $ and that each inserted variable is fresh.

\noindent{\small\parbox{.4\textwidth}{
  \begin{align*}
  &\textsc{InsertMetas}(El_s) = El_i\\
  &\textsc{InsertMetas}(U_s) = U_i
\end{align*}}
\parbox{.5\textwidth}{
  \begin{align*}
    &\textsc{InsertMetas}(u_s) = u_i    \\
    &\textsc{InsertMetas}(\pi_{s_1,s_2}) = \pi_{i,j}
\end{align*}}
\vspace{-1.5em}
\begin{align*}
  &\textsc{InsertMetas}(c) = c~i_1 ... i_k \text{ where }k = \textsc{Arity}(c) \text{ and } c \neq El_s,U_s,u_s,\pi_{s_1,s_2}\\
  &\textsc{InsertMetas}(M) = M \text{ if }M \text{ is a variable } x \text{ or } \Type \text{ or } \Kind\\
  &\textsc{InsertMetas}(\Pi x : A . B) =  \Pi x : \textsc{InsertMetas}(A). \textsc{InsertMetas}(B)\\
  &\textsc{InsertMetas}(\lambda x : A . M) = \lambda x : \textsc{InsertMetas}(A). \textsc{InsertMetas}(M)\\
  &\textsc{InsertMetas}(M N) = \textsc{InsertMetas}(M)~\textsc{InsertMetas}(N)
\end{align*}}%

\begin{example}
  By applying $ \textsc{InsertMetas}  $  to the type and body of $ thm_2 $  we get
{\small\begin{align*}
  &\textsc{InsertMetas}(El_*~(\pi_{\square,*}~P:u_*.P \leadsto_{*,*}P)) = El_{i_1}~(\pi_{i_2,i_3}~P:u_{i_4}.P \leadsto_{i_5,i_6} P)\\
  &\textsc{InsertMetas}(thm_1~(\pi_{\square,*}~P:u_*.P \leadsto_{*,*} P)~thm_1) = thm_1~i_7~(\pi_{i_8,i_9}~P:u_{i_{10}}. P \leadsto_{i_{11},i_{12}}P)~(thm_1~i_{13})
  \end{align*}}%
\end{example}
\begin{remark}
Note that because our first step is erasing the universes that appear in the terms, this translation is defined for all PTS local signatures, and not only those in \textbf{I}. Therefore, it can be applied to proofs coming from systems featuring much more complex universes hierarchies then \textbf{I}, such as the PTS underlying the type systems of \textsc{Coq} and \textsc{Matita}.  
\end{remark}

Once the fresh level variables are inserted, the next step is to compute the constraints between levels. To do this, we use an approach similar to \cite{typechecking-with-universes}  and define a bidirectional type checking/inference algorithm.

Figures \ref{cstr-conv} and \ref{cstr-type} define rules for computing constraints required for two terms to be convertible or for a term to be typable, respectively. In these rules, we write $ \hat{M} $ for the weak head normal form of $ M $ when it exists --- thus $\hat{(-)}$ is a partial function, but becomes total if we suppose the termination of $\red$. As usual, $ M \Rightarrow A $ denotes type inference, whereas $ M \Leftarrow A $ denotes type checking. We also write $ M \Rightarrow_{sort} \textbf{s} $ or $ M \Rightarrow_\Pi \Pi x : A. B$ as a shorthand for $ M \Rightarrow A' $ and $ \hat{A'}=\textbf{s} $ or $ \hat{A'}=\Pi x: A. B $ respectively.

\begin{figure}[ht]
  {\small
  \begin{center}
\end{center}
\begin{center}
  \AxiomC{$ l, l'~Level $}
\UnaryInfC{$l \teq^? l' \downarrow \{l = l'\}$}
\DisplayProof
\hskip 1.5em  
  \AxiomC{$ M = x, c, \Type, \Kind$}
  \UnaryInfC{$M \teq^? M \downarrow \emptyset$}
  \DisplayProof
\hskip 1.5em    
\AxiomC{$ M \teq^? M' \downarrow C_1$}
\AxiomC{$ \hat{N} \teq^? \hat{N'} \downarrow C_2$}
\BinaryInfC{$M N \teq^? M' N' \downarrow C_1 \cup C_2$}
\DisplayProof
\end{center}
\begin{center}
\AxiomC{$ \hat{A} \teq^? \hat{A'} \downarrow C_1$}
\AxiomC{$ \hat{B} \teq^? \hat{B'} \downarrow C_2$}
\BinaryInfC{$ \Pi x: A. B \teq^? \Pi x : A'. B' \downarrow C_1 \cup C_2$}
\DisplayProof
  \hskip 0.7em    
\AxiomC{$ \hat{A} \teq^? \hat{A'} \downarrow C_1$}
\AxiomC{$ \hat{M} \teq^? \hat{M'} \downarrow C_2$}
\BinaryInfC{$\lambda x : A. M \teq^? \lambda x : A'. M' \downarrow C_1 \cup C_2$}
\DisplayProof
\end{center}}
\vspace{-1em}
\caption{Inference rules for computing constraints for two terms in whnf to be convertible}
\label{cstr-conv}
\end{figure}

Intuitively, these judgments define a conditional typing relation that depends on the constraints  being satisfied. This intuition is formalized by the following results.

\begin{definition}
  Given a level substitution $ \theta $ (sending level variables to levels) and a set of constraints $ C $, containing pairs of levels $ l = l' $, we write $ \theta \vDash C $ when $ \text{for all } l = l' \in C ,l\theta \lvleq l' \theta $.
\end{definition}

\begin{lemma}
If $ M \teq^? N \downarrow C $ and $ \theta \vDash C $ then $ M \theta \teq N\theta $.
\end{lemma}
\tolong{\begin{proof}
By induction on $M \teq^{?} N \downarrow C$. All cases as similar, we do the application case as an example. By IH we have $M \theta \equiv M' \theta$ and $\hat{N}\theta \equiv \hat{N'} \theta$. Because $N \reds \hat{N}$ and $N' \reds \hat{N'}$, we also have $N\theta \reds \hat{N}\theta$ and $N'\theta \reds \hat{N'}\theta$, and thus $N \theta \equiv \hat{N}\theta \equiv \hat{N'}\theta \equiv N' \theta$. Therefore, $(M N)\theta = (M \theta) (N \theta) \equiv (M' \theta) (N' \theta)=(M'N')\theta$.
\end{proof}}

\begin{figure}
  {\footnotesize
\begin{center}
\AxiomC{$ c : A := M \in \Sigma_{UPP},\Delta \text{ or } c : A  \in \Sigma_{UPP},\Delta $}
\RightLabel{\textsc{Cons}}
\UnaryInfC{$\Sigma_{UPP},\Delta;\Gamma \vdash c \Rightarrow A \downarrow \emptyset $}
\DisplayProof
\hskip 1.5em
\AxiomC{$ x : A  \in \Gamma $}
\RightLabel{\textsc{Var}}
\UnaryInfC{$\Sigma_{UPP},\Delta;\Gamma \vdash x \Rightarrow A\downarrow \emptyset  $}
\DisplayProof
\end{center}
\begin{center}
\AxiomC{$i \in \mathcal{M}$}
\RightLabel{\textsc{Lvl-Var}}
\UnaryInfC{$\Sigma_{UPP},\Delta;\Gamma \vdash i \Rightarrow Level\downarrow \emptyset  $}
\DisplayProof
\hskip 1.5em
\AxiomC{}
\RightLabel{\textsc{Sort}}
\UnaryInfC{$\Sigma_{UPP},\Delta;\Gamma \vdash \Type \Rightarrow\Kind\downarrow \emptyset$}
\DisplayProof
\end{center}
\begin{center}
  \AxiomC{$\Sigma_{UPP},\Delta;\Gamma \vdash A \Leftarrow \Type \downarrow C_1$}
  \AxiomC{$\Sigma_{UPP},\Delta;\Gamma, x : A \vdash B \Rightarrow_{sort} \textbf{s} \downarrow C_2$}
\RightLabel{\textsc{Prod}}
\BinaryInfC{$\Sigma_{UPP},\Delta;\Gamma \vdash \Pi x : A. B \Rightarrow \textbf{s} \downarrow C_1 \cup C_2$}
\DisplayProof
\end{center}
\begin{center}
  \AxiomC{$\Sigma_{UPP},\Delta;\Gamma \vdash A \Leftarrow \Type \downarrow C_1$}
  \AxiomC{$\Sigma_{UPP},\Delta;\Gamma , x : A \vdash M \Rightarrow B \downarrow C_3$}
  \AxiomC{$\Sigma_{UPP},\Delta;\Gamma ,x : A \vdash B \Rightarrow_{sort} \textbf{s}\downarrow  C_2$}
  \RightLabel{\textsc{Abs}}
\TrinaryInfC{$\Sigma_{UPP},\Delta;\Gamma \vdash \lambda x : A. M \Rightarrow \Pi x : A. B \downarrow C_1 \cup C_2 \cup C_3$}
\DisplayProof
\end{center}
\begin{center}
  \AxiomC{$\Sigma_{UPP},\Delta;\Gamma \vdash M \Rightarrow_\Pi \Pi x: A.B \downarrow C_1$}
  \AxiomC{$\Sigma_{UPP},\Delta;\Gamma \vdash N \Leftarrow A \downarrow C_2$}
    \RightLabel{\textsc{App}}
\BinaryInfC{$\Sigma_{UPP},\Delta;\Gamma \vdash M N \Rightarrow B\{N/x\} \downarrow C_1 \cup C_2$}
\DisplayProof
\end{center}
\begin{center}
  \AxiomC{$\Sigma_{UPP},\Delta;\Gamma \vdash M \Rightarrow A \downarrow C_1$}
  \AxiomC{$\hat{A} \teq^? \hat{B} \downarrow C_2$}
  \RightLabel{\textsc{Check}}
\BinaryInfC{$\Sigma_{UPP},\Delta;\Gamma \vdash M \Leftarrow B \downarrow C_1 \cup C_2$}
\DisplayProof
\end{center}}
\vspace{-1em}
\caption{Inference rules for computing constraints for a term to be typable}
\label{cstr-type}
\end{figure}

Let us write $ \vec{i}_X $  for the free level variables in $ X $. We also shorten $ \vec{i} : Level $ as $ \vec{i} $.

\begin{lemma}\label{cstr-calc}
  Given a level substitution $ \theta $, suppose $ \Sigma_{UPP},\Delta;\vec{i}_{\Gamma\theta},\Gamma\theta~\WF $.
  \begin{itemize}
  \item If $\Sigma_{UPP},\Delta;\Gamma\vdash M \Rightarrow A \downarrow C $ and $ \theta \vDash C $  then $ \Sigma_{UPP},\Delta;\vec{i}_{\Gamma\theta}\cup\vec{i}_{M\theta}\cup\vec{i}_{A\theta},\Gamma\theta\vdash M\theta : A\theta $
  \item If $\Sigma_{UPP},\Delta;\Gamma\vdash M \Leftarrow A \downarrow C $, $ \theta \vDash C $ and $ \Sigma_{UPP},\Delta;\vec{i}_{\Gamma\theta}\cup\vec{i}_{A\theta},\Gamma\theta \vdash A\theta : \textbf{\textup{s}} $  then we have  $ \Sigma_{UPP},\Delta;\vec{i}_{\Gamma\theta}\cup\vec{i}_{M\theta}\cup\vec{i}_{A\theta},\Gamma\theta\vdash M\theta : A\theta $
  \end{itemize}
\end{lemma}
\tolong{\begin{proof}
  By induction on the derivation, the cases \textsc{Var}, \textsc{Cons}, \textsc{Lvl-Var}, and \textsc{Sort} being easy.

  \textbf{Case Prod}: We can show $\Sigma_{UPP},\Delta;\vec{i}_{\Gamma\theta},\Gamma\theta \vdash \Type : \Kind$, therefore by the IH on the first premise we get $\Sigma_{UPP},\Delta;\vec{i}_{\Gamma\theta} \cup \vec{i}_{A \theta},\Gamma\theta \vdash A \theta : \Type$. Using this, we can also derive $\Sigma_{UPP},\Delta;\vec{i}_{\Gamma\theta} \cup \vec{i}_{A \theta},\Gamma\theta, A\theta~\WF$. Therefore, by IH on the second premise we have  $\Sigma_{UPP},\Delta;\vec{i}_{\Gamma\theta} \cup \vec{i}_{A \theta}\cup\vec{i}_{B \theta}\cup\vec{i}_{T \theta},\Gamma\theta, A\theta \vdash B \theta : T \theta$, where $T \red^{*} \textbf{s}$. Therefore, we also have $T\theta \red^{*} \textbf{s}$, and thus we can derive $\Sigma_{UPP},\Delta;\vec{i}_{\Gamma\theta} \cup \vec{i}_{A \theta}\cup\vec{i}_{B \theta}\cup\vec{i}_{T \theta},\Gamma\theta, A\theta \vdash B \theta : \textbf{s}$. By applying the substitution lemma with the substitution sending all $i \in \vec{i}_{T\theta}\setminus (\vec{i}_{\Gamma\theta} \cup \vec{i}_{A \theta}\cup\vec{i}_{B \theta})$ to $\lzero$, we get $\Sigma_{UPP},\Delta;\vec{i}_{\Gamma\theta} \cup \vec{i}_{A \theta}\cup\vec{i}_{B \theta},\Gamma\theta, A\theta \vdash B \theta : \textbf{s}$. By weakening on $\Sigma_{UPP},\Delta;\vec{i}_{\Gamma\theta} \cup \vec{i}_{A \theta},\Gamma\theta \vdash A \theta : \Type$, we get $\Sigma_{UPP},\Delta;\vec{i}_{\Gamma\theta} \cup \vec{i}_{A \theta}\cup \vec{i}_{B \theta},\Gamma\theta \vdash A \theta : \Type$. Hence, it suffices to apply \texttt{Prod} to conclude $\Sigma_{UPP},\Delta;\vec{i}_{\Gamma\theta} \cup \vec{i}_{A \theta}\cup \vec{i}_{B \theta},\Gamma\theta \vdash \Pi x : A\theta. B \theta : \textbf{s}$.

  \textbf{Case Abs}: We can show $\Sigma_{UPP},\Delta;\vec{i}_{\Gamma\theta},\Gamma\theta \vdash \Type : \Kind$, therefore by the IH on the first premise we get $\Sigma_{UPP},\Delta;\vec{i}_{\Gamma\theta} \cup \vec{i}_{A \theta},\Gamma\theta \vdash A \theta : \Type$. Using this, we can also derive $\Sigma_{UPP},\Delta;\vec{i}_{\Gamma\theta} \cup \vec{i}_{A \theta},\Gamma\theta, A\theta~\WF$. Therefore, by IH on the second premise we have (1) $\Sigma_{UPP},\Delta;\vec{i}_{\Gamma\theta} \cup \vec{i}_{A \theta}\cup \vec{i}_{M\theta}\cup \vec{i}_{B\theta},\Gamma\theta, A\theta \vdash M \theta :B\theta$. By IH now on the third premise we have $\Sigma_{UPP},\Delta;\vec{i}_{\Gamma\theta} \cup \vec{i}_{A \theta}\cup\vec{i}_{B \theta}\cup\vec{i}_{T \theta},\Gamma\theta, A\theta \vdash B \theta : T \theta$, where $T \red^{*} \textbf{s}$. Therefore, we also have $T\theta \red^{*} \textbf{s}$, and thus we can derive $\Sigma_{UPP},\Delta;\vec{i}_{\Gamma\theta} \cup \vec{i}_{A \theta}\cup\vec{i}_{B \theta}\cup\vec{i}_{T \theta},\Gamma\theta, A\theta \vdash B \theta : \textbf{s}$. By applying the substitution lemma with the substitution sending all $i \in \vec{i}_{T\theta}\setminus (\vec{i}_{\Gamma\theta} \cup \vec{i}_{A \theta}\cup\vec{i}_{B \theta})$ to $\lzero$, we get $\Sigma_{UPP},\Delta;\vec{i}_{\Gamma\theta} \cup \vec{i}_{A \theta}\cup\vec{i}_{B \theta},\Gamma\theta, A\theta \vdash B \theta : \textbf{s}$. By applying weakening, we can derive  (2) $\Sigma_{UPP},\Delta;\vec{i}_{\Gamma\theta} \cup \vec{i}_{A \theta}\cup\vec{i}_{B \theta}\cup \vec{i}_{M\theta},\Gamma\theta, A\theta \vdash B \theta : \textbf{s}$. Using (1) and (2), we apply \texttt{Abs} to conclude $\Sigma_{UPP},\Delta;\vec{i}_{\Gamma\theta} \cup \vec{i}_{A \theta}\cup\vec{i}_{B \theta}\cup \vec{i}_{M\theta},\Gamma\theta \vdash \lambda x : A \theta. M\theta:\Pi x : A \theta. B \theta$.

  \textbf{Case App}: By IH on the first premise, we have $\Sigma_{UPP},\Delta;\vec{i}_{\Gamma\theta}\cup \vec{i}_{M\theta}\cup \vec{i}_{T\theta},\Gamma\theta \vdash M\theta : T \theta$, where $T \red^{*} \Pi x : A. B$. We thus also have $T \theta \reds \Pi x : A \theta. B \theta$, and thus by well-sortedness, subject reduction and the rule \texttt{Conv} we get $\Sigma_{UPP},\Delta;\vec{i}_{\Gamma\theta}\cup \vec{i}_{M\theta}\cup \vec{i}_{T\theta},\Gamma\theta \vdash M\theta : \Pi x :A \theta. B\theta$. By applying the substitution lemma with the substitution sending all $i \in \vec{i}_{T\theta}\setminus (\vec{i}_{\Gamma\theta}\cup\vec{i}_{M \theta}\cup\vec{i}_{A \theta}\cup\vec{i}_{B \theta})$ to $\lzero$, we get (1) $\Sigma_{UPP},\Delta;\vec{i}_{\Gamma\theta}\cup \vec{i}_{M\theta}\cup \vec{i}_{A\theta}\cup \vec{i}_{B\theta},\Gamma\theta \vdash M\theta : \Pi x :A \theta. B\theta$. By well-sortedness and inversion, we get $\Sigma_{UPP},\Delta;\vec{i}_{\Gamma\theta}\cup \vec{i}_{M\theta}\cup \vec{i}_{A\theta}\cup \vec{i}_{B\theta},\Gamma\theta \vdash A \theta : \Type$. By applying the substitution lemma with the substitution sending all $i \in (\vec{i}_{B\theta}\cup\vec{i}_{M\theta})\setminus (\vec{i}_{\Gamma \theta}\cup\vec{i}_{A \theta})$ to $\lzero$, we get $\Sigma_{UPP},\Delta;\vec{i}_{\Gamma\theta}\cup \vec{i}_{A\theta},\Gamma\theta \vdash A \theta : \Type$. Hence, we can apply the IH to the second premise, which gives (2) $\Sigma_{UPP},\Delta;\vec{i}_{\Gamma\theta}\cup \vec{i}_{N\theta}\cup \vec{i}_{A\theta},\Gamma\theta \vdash N \theta :A \theta$. By applying weakening to (1) and (2) we get $\Sigma_{UPP},\Delta;\vec{i}_{\Gamma\theta}\cup \vec{i}_{M\theta}\cup \vec{i}_{N\theta}\cup \vec{i}_{A\theta}\cup \vec{i}_{B\theta},\Gamma\theta \vdash M\theta : \Pi x :A \theta. B\theta$ and $\Sigma_{UPP},\Delta;\vec{i}_{\Gamma\theta}\cup \vec{i}_{M\theta}\cup \vec{i}_{N\theta}\cup \vec{i}_{A\theta}\cup \vec{i}_{B\theta},\Gamma\theta \vdash N \theta :A \theta$. By applying the rule \texttt{App} we get $\Sigma_{UPP},\Delta;\vec{i}_{\Gamma\theta}\cup \vec{i}_{M\theta}\cup \vec{i}_{N\theta}\cup \vec{i}_{A\theta}\cup \vec{i}_{B\theta},\Gamma\theta \vdash (M\theta) (N\theta) :  B\theta \{N\theta/x\}$. Finally, by applying the substitution lemma with the substitution sending all $i \in \vec{i}_{A\theta}\setminus (\vec{i}_{\Gamma \theta}\cup\vec{i}_{M \theta}\cup\vec{i}_{N \theta}\cup \vec{i}_{B \theta})$ to $\lzero$, we conclude $\Sigma_{UPP},\Delta;\vec{i}_{\Gamma\theta}\cup \vec{i}_{M\theta}\cup \vec{i}_{N\theta}\cup \vec{i}_{B\theta},\Gamma\theta \vdash (MN) \theta :  (B\{N/x\})\theta$.

\textbf{Case Check}: By hypothesis, we have (1) $\Sigma_{UPP},\Delta;\vec{i}_{\Gamma\theta} \cup \vec{i}_{B\theta}, \Gamma \theta\vdash B \theta : \textbf{s}$. By the IH applied to the first premise, we have  (2) $\Sigma_{UPP},\Delta;\vec{i}_{\Gamma\theta} \cup \vec{i}_{M\theta}\cup \vec{i}_{A\theta},\Gamma\theta\vdash M \theta : A \theta$. By applying weakening to (1) and (2) we get $\Sigma_{UPP},\Delta;\vec{i}_{\Gamma\theta}\cup \vec{i}_{M\theta} \cup \vec{i}_{A\theta}\cup \vec{i}_{B\theta},\Gamma\theta\vdash B \theta : \textbf{s}$ and $\Sigma_{UPP},\Delta;\vec{i}_{\Gamma\theta} \cup \vec{i}_{M\theta}\cup \vec{i}_{A\theta}\cup \vec{i}_{B\theta},\Gamma\theta\vdash M \theta : A \theta$.  By Lemma \ref{cstr-calc}, we have $ \hat{A}\theta \equiv\hat{B}\theta$, and thus $A \theta \equiv B \theta$. Hence, by the rule \texttt{Conv} we get $\Sigma_{UPP},\Delta;\vec{i}_{\Gamma\theta} \cup \vec{i}_{M\theta}\cup \vec{i}_{A\theta}\cup \vec{i}_{B\theta},\Gamma\theta\vdash M \theta : B \theta$. Finally, by applying the substitution lemma with the substitution sending all $i \in \vec{i}_{A\theta}\setminus (\vec{i}_{\Gamma \theta}\cup\vec{i}_{M \theta}\cup \vec{i}_{B \theta})$ to $\lzero$, we conclude $\Sigma_{UPP},\Delta;\vec{i}_{\Gamma\theta} \cup \vec{i}_{M\theta}\cup \vec{i}_{B\theta},\Gamma\theta\vdash M \theta : B \theta$.
\end{proof}}

\begin{example}
  We can use the rules  with our running example. We first calculate    $\Sigma_{UPP},\Delta_{thm_1};- \vdash El_{i_1}~(\pi_{i_2,i_3}~P:u_{i_4}.P \leadsto_{i_5,i_6}P)  \Rightarrow_{sort} \textbf{s} \downarrow C_1$. Therefore, any substitution $ \theta $ with $ \theta \vDash C_1 $  applied to the previous term results in a valid type. We then calculate
{\begin{align*}
\Sigma_{UPP},\Delta_{thm_1};- \vdash ~&thm_1~i_7~(\pi_{i_8,i_9}~P:u_{i_{10}}. P\leadsto_{i_{11},i_{12}}P)~(thm_1~i_{13}) \\ &\Leftarrow El_{i_1}~(\pi_{i_2,i_3}~P:u_{i_4}. P\leadsto_{i_5,i_6}P) \downarrow C_2
       \end{align*}}%
This gives $ C_1 \cup C_2 = \{i_8 = \lsucc~i_{10}, i_{11} = i_{10},i_{12}=i_{10},i_9=i_{10}\sqcup i_{12}, i_8 \sqcup i_9  = i_7, i_{13}=i_{10}, i_1 = i_2\sqcup i_3, \lsucc~i_4 = i_2, i_5 = i_4, i_6 = i_4, i_3 = i_5\sqcup i_6, i_4 = i_{10}\}$.
\end{example}

Once the constraints are computed the next step is to solve them. However, as explained in Section \ref{sec:firstlook}, we do not want a numerical assignment of level variables that satisfies the constraints, but rather a general symbolic solution which allows the term to be instantiated later at different universe levels. This leads us to use equational unification but, as levels are not purely syntactic entities, one needs to devise a unification algorithm for the equational theory $ \lvleq $. For now, let us postpone this to the next section and assume we are given a (partial) function $ \textsc{Unify} $ which computes from a set of constraints $ C $ a unifier $ \theta $ -- that is, a substitution satisfying $ \theta \vDash C $. We however do not assume that $ \theta $ is the most general unifier -- as we show later in Theorem \ref{no-mgu}, such a most general unifier might not  exist.

After an unifier $ \theta $ is found, the final step is then to apply it.
\begin{example}
  Given the previous computed constraints $ C_1 \cup C_2 $ we can compute the substitution $ \theta  $ which sends all variables to $ i_4 $, except for $ i_1, i_2, i_7,i_8  $, which are sent to $ \lsucc~i_4 $, and verify that $ \theta \vDash C_1\cup C_2 $. By applying $ \theta $, and by Lemma \ref{cstr-calc}, we have
  {\begin{align*}
\Sigma_{UPP},\Delta_{thm_1};i_4 \vdash ~&thm_1~(\lsucc~i_4)~(\pi_{(\lsucc~i_4),i_4}~P:u_{i_{4}}. P\leadsto_{i_{4},i_{4}}P)~(thm_1~i_{4}) \\ &: El_{(\lsucc~i_4)}~(\pi_{(\lsucc~i_4),i_4}~P:u_{i_4}. P\leadsto_{i_4,i_4}P)
         \end{align*}}
       Note that in this term, the constant $ thm_1 $ is used at two different universe levels.
\end{example}

\begin{figure}
  {\footnotesize
\begin{flalign*}
  |-|=&~ -\\
  |\Delta, c : A|=
      &\text{ let }A' = \textsc{InsertMetas}(A)\\
      &\text{ let }C \text{ be such that } \Sigma_{UPP},|\Delta|;-\vdash A' \Rightarrow_{sort}  \textbf{s} \downarrow C\text{ else }raise~\bot\text{ if no such }C\\
      &\text{ let }\theta = \textsc{Unify}(C)\text{ else }raise~\bot\text{ if no such }\theta\\
            &\text{ let }\vec{i} = \vec{i}_{A'\theta}\\
      &~|\Delta|, c : \Pi \vec{i} : Level. A'\theta\\
  |\Delta, c : A := M|=
      &\text{ let }A', M' = \textsc{InsertMetas}(A), \textsc{InsertMetas}(M)\\
      &\text{ let }C_1 \text{ be such that } \Sigma_{UPP},|\Delta|;-\vdash A' \Rightarrow_{sort} \textbf{s} \downarrow C_1 \text{ else }raise~\bot\text{ if no such }C_1\\
      &\text{ let }C_2 \text{ be such that } \Sigma_{UPP},|\Delta|;-\vdash M' \Leftarrow A' \downarrow C_2, \text{ else }raise~\bot\text{ if no such }C_2\\
      &\text{ let }\theta = \textsc{Unify}(C_1 \cup C_2)\text{ else }raise~\bot\text{ if no such }\theta\\
      &\text{ let }\tau = i \mapsto \text{if}~ i \in \vec{i}_{M'\theta}\setminus \vec{i}_{A'\theta} \text{ then }\lzero \text{ else }i \\
        &\text{ let }\vec{i} = \vec{i}_{A'\theta}\\
      &~|\Delta|, c : \Pi \vec{i} : Level. A'\theta := \lambda \vec{i} : Level.M'\theta\tau
\end{flalign*}}
\vspace{-2em}
\caption{Pseudocode of the predicativization algorithm}
\label{predicativize-alg}
\end{figure}

The final algorithm can now be described by the  pseudocode in Figure \ref{predicativize-alg}. The algorithm might fail at any point when either it is not able to compute the constraints, or if the unification algorithm is not capable of inferring a substitution from the constraints. However, if the algorithm returns, its correctness is guaranteed by the following theorem:

\begin{theorem}\label{algcorrect}
If $ |\Delta| $ is defined, then $ \Sigma_{UPP},|\Delta|;-~\WF$.
\end{theorem}
\begin{proof}
  By induction on $ \Delta $, the base case being trivial. For the induction step, we have either $ \Delta = \Delta'; c : A  $ or $ \Delta = \Delta'; c : A := M $. In both cases, if $ |\Delta| $ is defined, then so is $ |\Delta'| $, and thus by induction hypothesis we have $ |\Delta'|;-~\WF $. We proceed with a case analysis on the entry.

  \textbf{Definition:} As $ \Sigma_{UPP},|\Delta'|; -\vdash A'\Rightarrow T \downarrow C_1 $ and $ \theta \vDash C_1 $, by Lemma \ref{cstr-calc} we get $ \Sigma_{UPP},|\Delta'|;\vec{i}_{A'\theta}\cup\vec{i}_{T\theta} \vdash A'\theta:T\theta $. Because $ \hat{T}= \textbf{s} $, we also have $ T\theta \red^* \textbf{s} $, hence we can derive $ \Sigma_{UPP},|\Delta'|; \vec{i}_{A'\theta}\cup\vec{i}_{T\theta} \vdash A'\theta : \textbf{s} $. By applying the substitution lemma with the substitution sending every variable $ i $  in $ \vec{i}_{T\theta} $ but not in $ \vec{i}_{A'\theta}$ to $ \lzero $, we get $\Sigma_{UPP},|\Delta'|; \vec{i}_{A'\theta} \vdash A'\theta : \textbf{s}$. Because we also have $ \Sigma_{UPP},|\Delta'|;-\vdash M' \Leftarrow A' \downarrow C_2 $ and $ \theta \vDash C_2 $, by Lemma \ref{cstr-calc} again we get $ \Sigma_{UPP},|\Delta'|;\vec{i}_{M'\theta}\cup\vec{i}_{A'\theta}\vdash M'\theta:A'\theta $. By applying the substitution lemma with the substitution $\tau$, we get $\Sigma_{UPP},|\Delta'|;\vec{i}\vdash M'\theta\tau:A'\theta $, where $\vec{i}:= \vec{i}_{A'\theta}$.  Finally, by abstracting each free level variable, we get $ \Sigma_{UPP},|\Delta'|;-\vdash \lambda \vec{i} : Level.M'\theta\tau : \Pi\vec{i}: Level.A'\theta $. Hence, we can derive $ \Sigma_{UPP},|\Delta'|,c:\Pi \vec{i} : Level.A'\theta := \lambda \vec{i}:Level.M'\theta\tau;-~\WF$.

  \textbf{Constant:} Similar to the previous case.\qedhere
\end{proof}

\begin{remark}
One could also wonder if the algorithm always terminates and produces a result (be it a valid signature or $ \bot $). By supposing strong normalization for \textbf{UPP}, and by checking at each step of the rules in Figure \ref{cstr-type} that the constraints are consistent, one could show termination of the algorithm by using a similar technique as in \cite{typechecking-with-universes}. As we do not investigate strong normalization of \textbf{UPP} in this paper, we leave termination of the algorithm for future work. However, as we will see in Section \ref{sec:matita}, when using it in practice we were able to translate many proofs without non-termination issues.
\end{remark}

Our algorithm relies on an unspecified function $ \textsc{Unify} $ in order to solve the constraints. In order to fully specify it, we thus present an unification algorithm for $ \lvleq $ in the next section. As we will discuss, the unification algorithm we propose is not guaranteed to always find a most general unifier whenever there is one. However, note that our predicativization algorithm can in principle be used with any unification algorithm for $ \lvleq $. Therefore, if we have a better unification algorithm in the future, we do not have the modify the algorithm of Figure \ref{predicativize-alg}.

\section{Solving level constraints}
\label{sec:solving}

Before addressing the problem of how to solve level constraints, the first natural question that comes to mind is if one can always find a most general unifier (mgu) when the constraints are solvable. The following result answers this negatively.

\begin{theorem}\label{no-mgu}
Not all solvable problems of unification modulo $ \lvleq $  over levels have most general unifiers.
\end{theorem}

\begin{proof}
  Consider the equation $  \lsucc~i_1 = i_2 \sqcup i_3 $, which is a solvable unification problem, and suppose it  had a mgu $ \theta $. Note that $ \theta_1 = i_1 \mapsto \lzero, i_2 \mapsto \lsucc~\lzero, i_3 \mapsto \lzero $ is also  a solution, thus there is some $ \tau $ such that $ i_3\theta \tau \lvleq \lzero $. Therefore, there can be no occurrence of $ s $ in $ i_3\theta $. By taking $ \theta_2 = i_1 \mapsto \lzero, i_2 \mapsto \lzero, i_3 \mapsto \lsucc~\lzero $ we can show similarly that there can be no occurrence of $ \lsucc $ in $ i_2\theta $. But by taking the substitution $ \theta' = \_ \mapsto \lzero$ mapping all variables to $ \lzero $, we get $ (i_2 \sqcup i_3)\theta\theta' \lvleq \lzero $, which cannot be equivalent to $ (\lsucc~i_1)\theta\theta' $. Hence, $ \lsucc~i_1 = i_2 \sqcup i_3 $ has no mgu.
\end{proof}

Therefore, no unification algorithm for $ \lvleq $ can always produces a mgu. Hence, our algorithm will produce three kinds of results: either it produces a substitution, in which case it is a mgu; or it produces $ \bot $, in which case there is no solution to the constraints; or it gets stuck on a set of constraints that it cannot handle. Still, it is not guaranteed to compute a mgu whenever there is one.

Before presenting the algorithm, the first issue we have to address is the fact that levels can have multiple equivalent representations. It would be convenient if we had a syntactical way to compare them. Thankfully, previous works have already addressed this problem.

Let us assume from this point on that level variables in $ \mathcal{I} $ admit a total order $ < $. Given a strictly increasing sequence of level variables $ V = i_1,...,i_k $, and an $ V $-indexed family of levels $ \{l_i\}_{i \in V} $, let $ \sqcup_{i \in V}l_i $ denote the term $ l_1 \sqcup (l_2 \sqcup ... (l_{k-1} \sqcup l_k) ...) $. Moreover, given a natural number $ k $, let $ \lsucc^{k}~l $ be inductively defined by $ \lsucc^{0}~l = l $ and $ \lsucc^{n+1}~l = \lsucc~(\lsucc^n~l) $.

\begin{definition}[Level normal form]
A level is in normal form when it is  of the form $\lsucc^k~\lzero \sqcup (\sqcup_{i \in V} \lsucc^{n_i}~i) $ with $ n_i \leq k $ for all $ i $.
\end{definition}

Previous works \cite{guillaume,gaspard,blanqui22fscd} have established that for every level $ l $ there is a unique level in normal form, which we refer to as $ \hat{l} $,  with $ l \lvleq \hat{l} $ -- see for instance Lemma 6.2.5 of \cite{gaspard}. We will not describe explicitly here the algorithm for computing normal forms, as this has already been thoroughly explained in previous works, such as in \cite{guillaume,blanqui22fscd} -- we note  nevertheless that this procedure is sketched in the proof of Proposition \ref{semanticlvl}.

We also define a notion of normal form for constraints.

\begin{definition}
  A constraint $ l_1 = l_2 $ is said to be in normal form if
  \begin{enumerate}
  \item Both $ l_1, l_2 $ are in normal form -- so we write $ l_p =  \lsucc^{k_p}~\lzero \sqcup (\sqcup_{i \in V_p} \lsucc^{n^p_i}~i) $ for $ p = 1, 2 $
  \item If $ i \in V_1 \cap V_2 $, then $ n^1_i = n^2_i $
  \item At least one of the numbers in $ \{k_1, k_2\}\cup \{n^1_i\}_{i \in V_1}\cup \{n^2_i\}_{i \in V_2} $ is equal to $ 0 $
  \end{enumerate}
\end{definition}

Every constraint can be put in normal form, and for this we can use the algorithm in Figure \ref{normal-cstr}. From the second line on, we use $ k^p, n_i^p $ to refer to the indices in the normal forms  of $ l_1, l_2 $. Moreover, the pseudocode should be read imperatively, in the sense that $ l_1, l_2, V_1, V_2 $ are updated at each step.

\begin{figure}[ht]
{\footnotesize\begin{align*}
  &\text{let }l_1, l_2 = \hat{l_1}, \hat{l_2}\\
  &\text{for each } i \in V_1\cap V_2\\
  &\quad\text{if } n^1_i < n^2_i\text{ then remove } \lsucc^{n^1_i}~i \text{ from }l_1 \\
  &\quad\text{else if } n^1_i > n^2_i\text{ then remove } \lsucc^{n^2_i}~i \text{ from }l_2 \\
  &\text{substract the minimum value of the set } \{k_1, k_2\}\cup \{n^1_i\}_{i \in V_1}\cup \{n^2_i\}_{i \in V_2}\text{ from all of its elements}
\end{align*}}
\vspace{-2em}
\caption{Imperative algorithm for putting a constraint in normal form}
\label{normal-cstr}     
\end{figure}

\newcommand{\cstrnf}[1]{\widehat{#1}}

Given a set of  constraints $ C $, let $ \cstrnf{C} $ denote the result of putting all constraints of $ C $ in normal form by applying the algorithm of  Figure \ref{normal-cstr}.

\begin{lemma}\label{cstr-nf}
For all substitutions $ \theta $, we have $ \theta \vDash C $ iff $ \theta \vDash \cstrnf{C} $.
\end{lemma}
\tolong{\begin{proof}
  It suffices to show that for each step transforming $ l_1 = l_2 $ into $ l_1' = l_2' $, we have $ l_1\theta \lvleq l_2 \theta $ iff $ l_1'\theta \lvleq l_2'\theta $. For the first part, which transforms $ l_1 = l_2$ into $ \hat{l_1} = \hat{l_2} $, because we have $ l_1 \lvleq \hat{l_1}$, $ l_2 \lvleq \hat{l_2} $, we thus deduce $ l_1\theta \lvleq \hat{l_1}\theta$ and $ l_2\theta \lvleq \hat{l_2}\theta $, from which the result follows.

  For the second part, it suffices to show that for any $ l_1, l_2, l, n $ with $ n > 0 $, we have $(\star)$
$l_1 \sqcup l \lvleq l_2 \sqcup (\lsucc^n~l) \iff l_1 \lvleq l_2 \sqcup (\lsucc^n~l)$. Note that  for any $ p_1, p_2, m, q \in \mathbb{N} $ with $ q >0 $ we have $max\{p_1, m\} = max\{p_2, q + m\} \iff p_1 = max\{p_2, q + m\}$. Indeed, the direction $ \Leftarrow $ is clear, whereas for $ \Rightarrow $ if we had $ m = max\{p_2, q + m\} $ then we would have $ m > m $, contradiction.  We then can show $ (\star) $ by  applying this fact together with Lemma \ref{semanticlvl}.

For the third part, it suffices to note that  $ \lsucc^m~l \sqcup \lsucc^m~l' \lvleq \lsucc^m~(l \sqcup l')  $ and that $ \lsucc^m~l \lvleq \lsucc^m~l' $ iff $ l \lvleq l' $.
\end{proof}}

By putting constraints in normal form, we can help our unification algorithm to find a solution, as shown by the following example.
 
\begin{example}
Consider the constraint $ i\sqcup \lsucc~(i \sqcup \lsucc~j) = j \sqcup \lsucc~(\lsucc~i) $ -- which, as we will see, cannot be treated by our unification algorithm if it is not normalized first. By first computing the level normal form of each side, we get $ \lsucc^2~\lzero \sqcup \lsucc~i \sqcup \lsucc^2~j = \lsucc^2~\lzero \sqcup \lsucc^2~i \sqcup j $. As both variables appear in both sides, we remove from each of the sides the occurrence with the smaller index, giving $  \lsucc^2~\lzero \sqcup \lsucc^2~j = \lsucc^2~\lzero \sqcup \lsucc^2~i$. Finally, as the minimum among all indices is $ 2 $, we subtract this from all of them, giving $ \lzero \sqcup j = \lzero \sqcup i $ -- a constraint that can be treated by our unification algorithm.
\end{example}

Write $\mathcal{I}(l)$ for the level variables appearing in $l$. Given a substitution $ \theta $, we define the sets $dom~\theta = \{i \mid i \neq i \theta\}$ and $range~\theta = \cup_{i\in dom~\theta} \mathcal{I}(i\theta)$, and the substitutions $ \widehat{\theta} = \{i \mapsto \widehat{i\theta}\}_{i \in dom~\theta} $, and $ \theta\{l/j\} = \{i  \mapsto i\theta\{l/j\}\}_{i \in dom~\theta}$. Finally, we also define $\mathcal{I}(\theta) = range~\theta\cup dom~\theta$.

\begin{figure}[ht]
{\footnotesize\begin{flalign*}
 &(\textbf{Trivial}) &\{l = l\} \cup C ; \theta &\leadsto C ; \theta\\
 &(\textbf{Orient}) & \{l = l'\} \cup C ; \theta &\leadsto \{l' = l\} \cup C ; \theta&\text{if }l' = \lzero \text{ or }\lzero \sqcup i \\
  &(\textbf{Eliminate 1}) &\{\lzero \sqcup i = l\} \cup C ; \theta &\leadsto \cstrnf{C\{l/i\}} ; \widehat{\theta\{l/i\}}, i \mapsto l &\text{if }i \notin l\\
&(\textbf{Eliminate 2})   &\{\lzero \sqcup i = l\} \cup C ; \theta &\leadsto &\text{if }\lsucc^m~i \in l\text{ with }m = 0\\
& & & \hspace{-6em}\text{let }l' = l \{i' / i\}  \text{ in }\cstrnf{C\{l'/i\}} ; \widehat{\theta\{l'/i\}}, i \mapsto l' & \text{for some } i'\in \mathcal{I}_{fresh}\setminus \mathcal{I}(i, l,C, \theta) \\
  &(\textbf{Decompose}) &\{\lzero = \lzero \sqcup (\sqcup_{i \in V}~i)  \} \cup C ; \theta &\leadsto \{\lzero \sqcup i = \lzero\}_{i \in V} \cup C; \theta &\\
&(\textbf{Clash}) &\{\lzero = l  \} \cup C ; \theta  &\leadsto \bot &\hspace{-1em}\text{ if } \lsucc^{n}~i \in l\text{ or } \lsucc^{n} ~\lzero \in l \text{ with } n\neq 0
       \end{flalign*}}
\vspace{-2em}
\caption{Unification algorithm for $ \lvleq $}
\label{unification}     
\end{figure}

We are now ready to present the unification algorithm, whose rules are given in Figure \ref{unification}. Steps are represented by rules of the form $ C; \theta \leadsto C'; \theta' $, with the pre-conditions that constraints in $ C $ are in normal form, $dom~\theta $ is disjoint from $ range~\theta$, the image of $ \theta $  contains only levels in normal form, and $dom~\theta$ is  disjoint from $ \mathcal{I}(C) $ --  these properties are preserved by each step. In rule (Eliminate 2), $\mathcal{I}_{fresh}\subsetneq \mathcal{I}$ is an infinite set of fresh level variables. Finally, it may happen that, for some non-empty sets of constraints $C$, no rule applies. This corresponds to the cases in which our algorithm gets stuck and does not produce a solution.

Let us write $ \theta_1\subseteq\theta_2 $ when for all $ i \in dom~\theta_1 $, $ i\theta_1  = i\theta_2  $ and $dom~\theta_{2} \setminus (dom~\theta_{1})\subseteq \mathcal{I}_{fresh}$ -- that is, $\theta_{2}$ extends $\theta_{1}$ only inside $\mathcal{I}_{fresh}$.

The following lemma is key in showing the main properties of our algorithm.

\begin{lemma}[Key lemma]\label{key}
  Suppose  $ C;\theta \leadsto C';\theta' $.  For all $ \tau $, if $ \tau \vDash C $ and $ \tau \lvleq \tau\circ\theta$ then there is a substitution $ \tau' $ with $ \tau \subseteq \tau' $ such that (1) $ \tau' \vDash C' $ and (2) $ \tau' \lvleq \tau'\circ\theta' $.
  Conversely, for all $ \tau $, if $ \tau \lvleq \tau \circ \theta' $ and $ \tau \vDash  C'$, then $ \tau \lvleq \tau \circ \theta $ and $ \tau \vDash C $.
\end{lemma}
\tolong{\begin{proof}
We first start with the following auxiliary lemma:

\begin{lemma}\label{aux}
  Suppose $ i \theta \lvleq l\theta $. Then we have
  \begin{enumerate}
  \item $ l'\theta \lvleq l' \{l/i\}\theta $, for all $ l' $
  \item $ \theta \vDash C $ iff $ \theta \vDash C\{l/i\} $.
  \end{enumerate}
\end{lemma}
\begin{claimproof}
Part (1) can be shown by a simple induction on $ l' $. For part (2), given $ l_1 = l_2 \in C $, we have $ l_1 \theta \lvleq l_1 \{l/i\}\theta$ and $ l_2 \theta\lvleq l_2 \{l/i\}\theta $, and thus $ l_1 \theta\lvleq l_2 \theta$ iff $ l_1 \{l/i\}\theta \lvleq l_2 \{l/i\}\theta $.
\end{claimproof}

We now continue with the proof of Lemma \ref{key}. It is done by case analysis on the rule, the cases \textbf{Trivial} and \textbf{Orient} being trivial. In the following, we might use Lemmas \ref{aux} and \ref{cstr-nf} implicitly.

  \textbf{Eliminate 1}: Suppose $ \tau \vDash \{\lzero \sqcup i = l\}\cup C $ and $ \tau \lvleq \tau \circ \theta $, and pose $ \tau' = \tau $. First note that we have $ i\tau \lvleq l\tau $, and thus $ i\tau' \lvleq l\tau' $.
  \begin{enumerate}
  \item We have $ \tau \vDash C $, but because $ i\tau \lvleq l\tau $ we get $ \tau \vDash C\{l/i\} $ and thus $ \tau \vDash \cstrnf{C\{l/i\}} $. Because $\tau'=\tau$, this shows (1).

  \item It suffices to show $j\tau' \simeq j \theta' \tau'$ for $j \in dom~\theta' = dom~\theta \cup \{i\}$. For all $ j \in dom~\theta $, we have $ j\tau'\lvleq j\theta\tau' \lvleq j\theta\{l/i\}\tau'   \lvleq \widehat{j\theta\{l/i\}}\tau' = j \theta'\tau'$, showing $j\tau' \simeq j\theta' \tau'$ for $j \in dom~\theta$. For $j = i$, this follows from the fact that $i\tau' \simeq l \tau'$  and $i\theta'=l$.
\end{enumerate}
  Conversely, suppose now $ \tau \vDash \cstrnf{C\{l/i\}} $ and $  \tau \simeq \tau \circ \theta'$.  As $ i\theta' = l $, it follows that $ i\tau \lvleq l \tau $, showing $ (\lzero \sqcup i)\tau \lvleq l\tau $. Moreover, $ \tau\vDash\cstrnf{C\{l/i\}} $ implies $ \tau \vDash C\{l/i\} $, which then implies $ \tau \vDash C $. Finally, for $ j \in dom~\theta $, $ j\tau  \lvleq j\theta'\tau = \widehat{j\theta\{l/i\}} \tau\lvleq j\theta\{l/i\} \tau \simeq j\theta\tau $, showing $\tau \simeq \tau \circ \theta$.

  \textbf{Eliminate 2}: Suppose $ \tau \vDash \{\lzero \sqcup i = l\}\cup C $ and $ \tau \lvleq \tau \circ \theta $.  Pose $ \tau' = \tau, i' \mapsto i\tau $. First note that $l' \tau' = l\{i'/i\}\tau' \simeq l \tau' \simeq (\lzero \sqcup i)\tau'\simeq i\tau'$, and thus $ i\tau' \lvleq l'\tau' $.
  \begin{enumerate}

    \item We have $ \tau \vDash C $ and thus $\tau' \vDash C$, but because $ i\tau' \lvleq l'\tau' $ we get $ \tau' \vDash C\{l'/i\} $ and thus $ \tau' \vDash \cstrnf{C\{l'/i\}} $.

  \item It suffices to show $j\tau' \simeq j \theta' \tau'$ for $j \in dom~\theta' = dom~\theta \cup \{i\}$. For all $ j \in dom~\theta $, we have $ j\tau'\lvleq j\theta\tau' \lvleq j\theta\{l'/i\}\tau'   \lvleq \widehat{j\theta\{l'/i\}}\tau' = j \theta'\tau'$, showing $j\tau' \simeq j\theta' \tau'$ for $j \in dom~\theta$. For $j = i$, this follows from the fact that $i\tau' \lvleq l' \tau'$ and $i\theta'=l'$.
  \end{enumerate}

  Conversely, suppose now $ \tau \vDash \cstrnf{C\{l'/i\}} $ and $  \tau = \tau \circ \theta'$. Let us first show that $(\lzero \sqcup i)\tau \simeq l \tau$. We can decompose $l$ as $l_{0} \sqcup i \simeq l$, where $l_{0}$ does not contain $i$.   Because $ i \tau \simeq i \theta'\tau$, we have $  i \tau \simeq l' \tau \simeq l_{0}\tau \sqcup i' \tau $, and thus $(\lzero \sqcup i)\tau \simeq l' \tau \simeq l_{0}\tau \sqcup i'\tau \simeq l_{0}\tau \sqcup l_{0}\tau \sqcup i'\tau \simeq l_{0}\tau \sqcup l' \tau \simeq l_{0}\tau \sqcup i \tau \simeq l \tau $, showing $(\lzero \sqcup i)\tau \simeq l \tau$. To conclude $\tau \vDash \{\lzero \sqcup i = l\}\cup C$ it now suffices to show $\tau \vDash C$, but   $ \tau\vDash\cstrnf{C\{l'/i\}} $ implies $ \tau \vDash C\{l'/i\} $, which then implies $ \tau \vDash C $. Finally, for $ j \in dom~\theta $, $ j\tau  \lvleq j\theta'\tau = \widehat{j\theta\{l'/i\}} \tau\lvleq j\theta\{l'/i\} \tau \simeq j\theta\tau $, showing $\tau \simeq \tau \circ \theta$.

  \textbf{Decompose}: Suppose $ \tau \vDash \{\lzero = \lzero \sqcup (\sqcup_{i \in V}i)  \} \cup C $ and $ \tau \lvleq \tau \circ \theta $.  Pose $ \tau' = \tau $. Point (2) is trivial. For (1), as $ \lzero \lvleq \sqcup_{i \in V} i\tau $, we must have $ i\tau \lvleq \lzero $ for all $ i \in V $. Thus $ (\lzero \sqcup i)\tau' \lvleq \lzero \tau'$ for all $ i \in V $. Finally, for all $ l_1 = l_2 \in C $, we have $ l_1\tau' \lvleq l_2 \tau' $ by hypothesis. A symmetric reasoning shows also the converse statement.
\end{proof}}

It is clear that  $ \leadsto$ does not always terminate, given that  some rules create constraints and in particular that the rule (Orient) can  loop. However, it is easy to check that these created constraints can always be eliminated by applying other rules. Indeed, the constraints created by (Decompose) can be eliminated by using (Eliminate 1), and the constraint created by (Orient) can be eliminated either by (Eliminate 1), (Eliminate 2), (Clash) or (Decompose), whose created constraints are eliminated once again using (Eliminate 1). Let $\leadsto_{0}$ be the relation that packs all of this into a single reduction, which therefore never creates constraints.

\begin{lemma}\label{termination}
$ \leadsto_0 $ terminates.
\end{lemma}
\tolong{\begin{proof}
Each step of $ \leadsto_0 $ decreases the number of constraints in $C$.
\end{proof}}

In the following theorems, we suppose  that $\mathcal{I}(C) $ and $ \mathcal{I}_{fresh}$ are disjoint.

\begin{theorem}\label{nosol}
If $ C; id \leadsto_0^* \bot $, then for no $ \theta $ we have $ \theta \vDash C $.
\end{theorem}
\tolong{\begin{proof}
If $ C; id \leadsto_0^* \bot $, then the calculation finishes with (Clash). If $ \theta \vDash C $, then by iterating Lemma \ref{key} we get that for some $ \theta' $, $ \lzero\theta' \lvleq (\lsucc^{k}~\lzero \sqcup (\sqcup_{i \in V}\lsucc^{n_i}~i))\theta'  $, where $k>0$ or $n_{i}>0$ for some $i$. But for any $ \sigma $ we have $ \trans{\lzero\theta'}_\sigma = 0 $ and $ \trans{(\lsucc^{k}~\lzero \sqcup (\sqcup_{i \in V}\lsucc^{n_i}~i))\theta'}_\sigma > 0 $, contradiction.
\end{proof}}

\begin{theorem}
If $ C; id \leadsto_0^* \emptyset; \theta $, then $ \theta $ is a most general unifier.
\end{theorem}
\begin{proof}
  Let $ \tau $ be a unifier. We thus have $ \tau \vDash C$. Moreover, we have $ \tau \lvleq \tau \circ id $. By iterating Lemma \ref{key}, we get a substitution $ \tau' $ such that $ \tau' \lvleq \tau' \circ \theta $ and $\tau \subseteq \tau'$. Because $dom~\tau' \setminus (dom~\tau) \subseteq \mathcal{I}_{fresh}$, which is disjoint from $\mathcal{I}(C)$, we have $ i\tau = i\tau'  $ for $ i \in \mathcal{I}(C) $. Hence, for $ i \in \mathcal{I}(C) $ we have $ i\tau \lvleq i\theta\tau' $, showing that $ \tau $ is an instance of $ \theta $.

  To show that $ \theta $ is a unifier,  note that  $ \theta = \theta \circ \theta $ and $ \theta \vDash \emptyset $, hence by iterating Lemma \ref{key} in the inverse direction we get $ \theta \vDash C $.
\end{proof}

We have seen that when the algorithm finishes with $ \emptyset;\theta $, then $ \theta $ is a mgu, and when it finishes with $ \bot $, then there is no solution to the constraints. However, the algorithm can also get stuck on constraints that it does not know how to solve.  In practice, it is very unsatisfying for the unification to get stuck, as this means that the whole predicativization algorithm has to halt. Thus, in order to prevent this, in our implementation we extended the unification with heuristics that are \textit{only} applied when none of the presented rules applies. Then, whenever the heuristics are applied, the universe polymorphic definition or declaration that is produced might not be the most general one.

\section{\textsc{Predicativize}, the implementation}
\label{sec:predicativize}

In this section we present \textsc{Predicativize}, an implementation of our algorithm. It is publicly available at \url{https://github.com/Deducteam/predicativize/}.

Our tool is implemented on top of \DkCheck{} \cite{saillard15phd}, a type-checker for \Dedukti{}, and thus does not rely neither on the codebase of \Agda{}, nor on the codebase of any other proof assistant. Like \Universo{} \cite{thire}, our implementation instruments \DkCheck{}'s conversion checking in order to implement the constraint computation algorithm described in Section \ref{sec:alg}.

Because the currently available type-checkers for \Dedukti{} do not implement rewriting modulo for equational theories other than AC (associative commutative), we used Genestier's encoding of levels \cite{guillaume} in order to  define the theory \textbf{UPP} in a \DkCheck{} file.

To see how everything works in practice, we invite the reader to download the code and run \texttt{make running-example}, which translates our running example and produces  a \Dedukti{} file \texttt{output/running_example.dk}  and an \Agda{} file \texttt{agda_output/running-example.agda}. In order to test the tool with a more realistic example, the reader can also run \texttt{make test_agda}, which translates a proof of Fermat's little theorem from the \Dedukti{} encoding of \textsc{HOL} \cite{sttfa} to \UPP{}.

In the following, let us go through some important particularities of how the tool works.

\vspace{-1em}
\paragraph*{User added constraints}

As we have seen, our tool tries to compute the most general type for a definition or declaration to be typable. However, it is not always desirable to have the most general type, as shown by the following example.

\begin{example}\label{succ}
  Consider the local signature
  {\small\begin{align*}
 \Delta = Nat : U_\square; zero : El_\square~Nat; succ : El_\square~(Nat\leadsto_{\square,\square}Nat)           
\end{align*}}%
 defining the natural numbers in \textbf{I}. The translation of this signature by our algorithm is
{\small\begin{align*}
 |\Delta| = Nat : \Pi i : Level. U_i; zero : \Pi i: Level.El_i~(Nat~i); succ :\Pi i~j : Level. El_{(i \sqcup j)}~((Nat~i)\leadsto_{i,j} (Nat~j)) 
\end{align*}}%
However, we normally would like to impose $ i $ to be equal to $ j $ in the type of $ succ $, or even to impose $ Nat $ not to be universe polymorphic.
\end{example}

In order to solve this problem, we added to \textsc{Predicativize} the possibility of adding constraints by the user, in such a way that we can for instance impose $ Nat $ to be in $ U_\lzero $, or $ i = j $ in the type of the successor. Adding constraints can also be useful to help the unification algorithm.

\vspace{-1em}
\paragraph*{Rewrite rules}
\label{subsec:rewrite}

The algorithm that we presented and proved correct covers two types of entries: definitions and constants. This is enough for translating proofs written in higher-order logic or similar systems, in which every step either poses an axiom or makes a definition or proof.

However, when dealing with full-fledged type theories, such as those implemented by \textsc{Coq} or \textsc{Matita}, which also feature inductive types, it is customary to use rewrite rules to encode recursion and pattern matching. If we simply ignore these rules when performing the translation, we would run into problems as the entries that appear after may need those rewrite rules to typecheck.

Therefore, our implementation  extends the presented algorithm and also translate rewrite rules. In order to do this, we use \DkCheck{}'s subject reduction checker to generate constraints and proceed similarly as in the algorithm. Because this feature is still work in progress, this step can require user intervention in some cases. In this case, the user has to manually add constraints over some symbols to help the translation.

\vspace{-1em}
\paragraph*{Agda output}
\label{subsec:agda}

\Predicativize{} produces files in \UPP{}, which is a subsystem of the encoding of \textsc{Agda}. In order to translate these files to \Agda{} itself, we also integrated in \Predicativize{} a translator that performs a simple syntactical translation from the \Agda{} encoding in \Dedukti{} to \Agda{}. For instance, \texttt{make test\_agda\_with\_typecheck} translates Fermat's Little Theorem proof from HOL to \Agda{} and typechecks it.

\section{Translating Matita's arithmetic library to Agda}
\label{sec:matita}

\begin{figure}[ht]
\begin{center}
  \includegraphics[width=\textwidth]{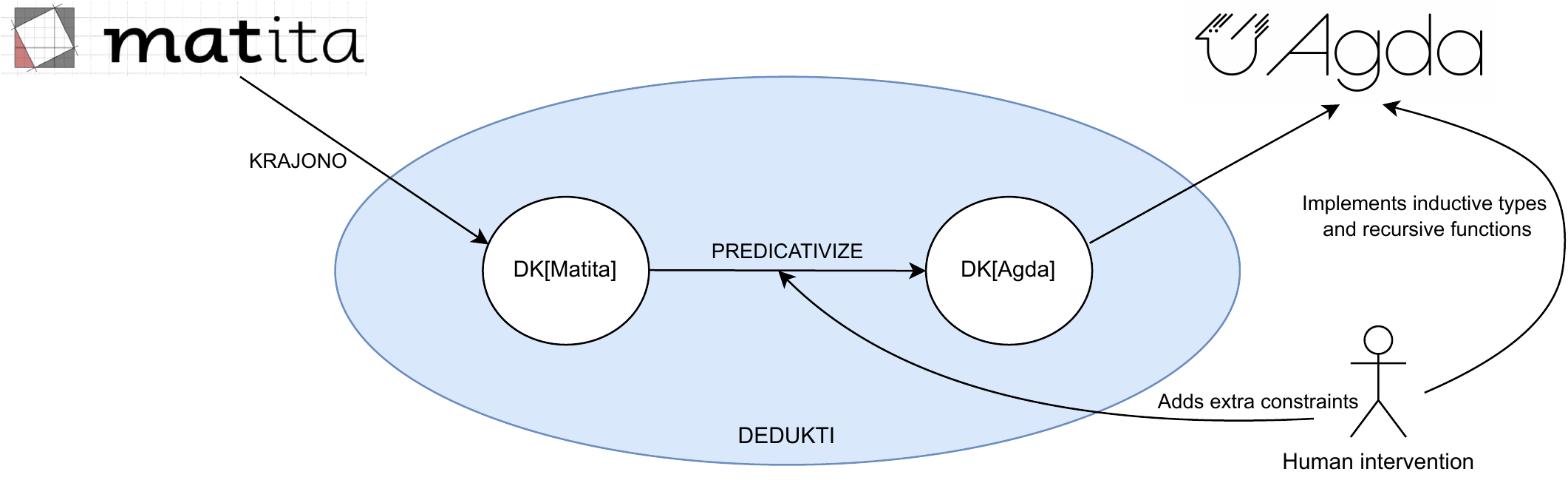}
\end{center}
\caption{Diagram representing the translation of \textsc{Matita}'s arithmetic library into \textsc{Agda}}
\label{matita}
\end{figure}

We now discuss how we used \textsc{Predicativize} to translate  \textsc{Matita}'s arithmetic library to \Agda{}. The translation is summarized in Figure \ref{matita}, where \textbf{DK}[X] stands for the encoding of system X in \Dedukti{}.

\textsc{Matita}'s arithmetic library was already available in \Dedukti{} thanks to  \textsc{Krajono}  \cite{assaf}, a translator from \textsc{Matita} to the encoding \textbf{DK}[Matita] in \Dedukti{}. Therefore, the first step of the translation was already done for us.

Then, using \textsc{Predicativize} we translated the library from \textbf{DK}[Matita] to \textbf{DK}[Agda] (which is a supersystem of \textbf{UPP}). As the encoding of \textsc{Matita}'s recursive functions uses rewrite rules, their translation required some user intervention to add constraints over certain symbols, as mentioned in the previous section. Once this step is done, the library is known to be predicative, as it typechecks in \textbf{DK}[Agda].

We then used \textsc{Predicativize} to translate these files to \Agda{} files. However, because the rewrite rules in the \Dedukti{} encoding cannot be translated to \Agda{}, and given that they are needed for typechecking the proofs, the library does not typecheck directly.

Therefore, to finish our translation we had to  define the inductive types and recursive functions manually in \Agda{}. This  step of our translation admittedly requires some time, however the effort is orders of magnitude less than rewriting the whole library in \Agda{}, specially given that the great majority of the library is made of proofs, whose translations we did not need to change. Note that this manual step is not exclusive to our work as it is also needed in \cite{sttfa}.

Defining inductive types also required us to add constraints. For instance, we saw in Example \ref{succ} that the successor symbol is translated as $ succ :\Pi i~j : Level. El_{(i \sqcup j)}~((Nat~i)\leadsto_{i,j} (Nat~j))  $, but in order to be able to implement this symbol as a constructor of an inductive type, we need to impose $ i = j $. If one then wishes to align $ Nat $ with the built-in type of natural numbers in \textsc{Agda}, we would also have to impose $ i = \lzero $, which would then allow us to replace $ Nat $ by the built-in type in the result of the translation.

The result of this translation is available at \url{https://github.com/thiagofelicissimo/matita_lib_in_agda} and, as far as we know, contains the very first proofs in \Agda{} of Bertrand's Postulate and Fermat's Little Theorem. It also contains a variety of other interesting results such as the Binomial Law, the Chinese Remainder Theorem, and the Pigeonhole Principle. Moreover, this library typechecks with \Agda{}'s \texttt{---safe} flag, attesting that it does not use any unsafe features.

\section{Conclusion}
\label{sec:conc}

We have tackled the problem of sharing proofs with predicative systems, by proposing an algorithm for it. Our implementation allowed to translate many non-trivial proofs from \textsc{Matita}'s arithmetic library to \Agda{}, showing that our algorithm  works well in practice.

Our solution uses unification modulo arithmetic equivalence on universe levels. We  designed an incomplete algorithm for this problem which is powerful enough for our needs. Still, one can wonder if there is an algorithm which always finds a most general solution when there is one, or if this problem is undecidable. One could  improve our algorithm by using ACUI unification to solve constraints not containing $\lsucc$. However, there are problems with most general unifiers that would also not be handled by this extension. \Agda{} also features an algorithm for solving level metavariables which uses an approach different from ours, but it does not seem to have been formalized in the literature. Therefore, the question if such  an algorithm exists seems to be open.

For future work, we would also like to look  at possible ways of making \textsc{Predicativize} less dependent on user intervention. In particular, the translation of inductive types and recursive functions involves some considerable manual work. Thus if we want to be able to translate larger libraries, there is definitely a need for automating this step.

\bibliography{ref}

\toshort{\appendix

\section{Typing rules for Dedukti and basic metaproperties}
\label{sec:dktyping}
\begin{figure}[h]
{\small
\begin{center}
  \AxiomC{}
\RightLabel{\texttt{Empty}}
\UnaryInfC{$-; -~\texttt{well-formed}$}
\DisplayProof
\hskip 1.5em
\AxiomC{$\Sigma;- \vdash A : \textbf{s}$}
\RightLabel{\texttt{Decl-cons}}
\LeftLabel{$c \notin \Sigma$}
\UnaryInfC{$\Sigma, c : A;-~\texttt{well-formed}$}
\DisplayProof
\end{center}
\begin{center}
\AxiomC{$\Sigma;- \vdash M : A$}
\RightLabel{\texttt{Decl-def}}
\LeftLabel{$c \notin \Sigma$}
\UnaryInfC{$\Sigma, c : A := M;-~\texttt{well-formed} $}
\DisplayProof
\hskip 1.5em
\AxiomC{$\Sigma;\Gamma \vdash A : \Type$}
\RightLabel{\texttt{Decl-var}}
\LeftLabel{$x \notin \Gamma$}
\UnaryInfC{$\Sigma;\Gamma, x : A~\texttt{well-formed} $}
\DisplayProof
\end{center}
\begin{center}
\AxiomC{$\Sigma;\Gamma~\texttt{well-formed}$}  
\RightLabel{\texttt{Cons}}
\LeftLabel{$c : A\text{ or }c : A := M \in \Sigma$}
\UnaryInfC{$\Sigma;\Gamma \vdash c : A $}
\DisplayProof
\hskip 1.5em
\AxiomC{$\Sigma;\Gamma~\texttt{well-formed}$}
\RightLabel{\texttt{Var}}
\LeftLabel{$ x : A \in \Gamma $}
\UnaryInfC{$\Sigma;\Gamma \vdash x : A  $}
\DisplayProof
\end{center}
\begin{center}
\AxiomC{$ \Sigma;\Gamma~\texttt{well-formed} $}
\RightLabel{\texttt{Sort}}
\UnaryInfC{$\Sigma;\Gamma \vdash \Type : \Kind$}
\DisplayProof
\hskip 1.5em
  \AxiomC{$\Sigma;\Gamma \vdash M : A $}
  \AxiomC{$\Sigma;\Gamma \vdash B : \textbf{s} $}  
  \RightLabel{\texttt{Conv}}
  \LeftLabel{$A \equiv B$}
\BinaryInfC{$\Sigma;\Gamma \vdash M : B $}
\DisplayProof
\end{center}
\begin{center} 
  \AxiomC{$\Sigma; \Gamma, x : A \vdash B : \textbf{s} $}  
\RightLabel{\texttt{Prod}}
\UnaryInfC{$\Sigma;\Gamma \vdash \Pi    x : A . B : \textbf{s}  $}
\DisplayProof
\hskip 1.5em
\AxiomC{$\Sigma;\Gamma \vdash M : \Pi x : A . B  $}
\AxiomC{$\Sigma; \Gamma \vdash N : A $}
\RightLabel{\texttt{App}}
\BinaryInfC{$\Sigma;\Gamma \vdash M N : B\{N/x\} $}
\DisplayProof
\end{center}
\begin{center}
  \AxiomC{$\Sigma; \Gamma, x : A \vdash B : \textbf{s} $}  
  \AxiomC{$\Sigma; \Gamma, x : A \vdash M : B $}
  \RightLabel{\texttt{Abs}}
\BinaryInfC{$\Sigma;\Gamma \vdash \lambda x : A . M :\Pi x : A . B  $}
\DisplayProof
\end{center}}
\caption{Typing rules for \Dedukti}
\label{typing-dk}
\end{figure}

We recall the following basic metaproperties of \Dedukti{}. Proofs can be found in \cite{frederic-phd, saillard15phd}. 

\begin{theorem}[Basic metaproperties]
~
  \begin{enumerate}
  \item Weakening:   If $ \Sigma;\Gamma \vdash M : A $, $ \Gamma \subseteq \Gamma' $ and $ \Sigma;\Gamma'~\WF $ then $ \Sigma;\Gamma' \vdash M : A $
  \item Substitution Lemma: If $ \Sigma; \Gamma,x:B,\Gamma' \vdash M : A $ and $ \Sigma;\Gamma \vdash N : B $ then $ \Sigma;\Gamma,\Gamma'\{N/x\}\vdash M\{N/x\} : A\{N/x\} $
  \item Well-sortedness: If $ \Sigma;\Gamma \vdash M : A $ then either $ A = \Kind $ or $ \Sigma;\Gamma \vdash A : \textup{\textbf{s}} $ for $ \textbf{\textup{s}}  = \Type$ or $ \Kind $.
  \item Subject reduction of $ \delta $: If $ \Sigma;\Gamma \vdash M : A $ and $ M \red_\delta M' $ then $ \Sigma;\Gamma \vdash M' : A $
  \item Subject reduction of $ \beta $: If injectivity of dependent product holds, then $ \Sigma;\Gamma \vdash M : A $ and $ M \red_\beta M' $ implies $ \Sigma;\Gamma \vdash M' : A $.
  \item Contexts are well typed: If $ x : A \in \Gamma $ then $ \Sigma;\Gamma \vdash A : \Type $
  \item Signatures are well typed: If $ c : A \in \Sigma$  then $ \Sigma;- \vdash A : \textup{\textbf{s}}$ and if $ c : A := M \in \Sigma $ then $  \Sigma;-\vdash M : A$
  \item Inversion of typing: Suppose $ \Sigma;\Gamma \vdash M : A $
    \begin{itemize}
    \item If $ M = x $ then $ x : A' \in \Gamma $ and $ A \equiv A' $
    \item If $ M = c $ then $ c : A'  \in \Sigma $ and $ A \equiv A' $
    \item If $ M = \Type $ then $ A \equiv \Kind $
    \item  $ M= \Kind $ is impossible
    \item If $ M = \Pi x : A_1. A_2 $ then $ \Sigma; \Gamma,x:A_1 \vdash A_2 : \textbf{\textup{s}} $ and $ \textbf{\textup{s}} \equiv A $
    \item If $ M = M_1 M_2 $ then $ \Sigma; \Gamma \vdash M_1 : \Pi x: A_1.A_2 $, $ \Sigma;\Gamma \vdash M_2 : A_1 $ and $ A_2\{M_2/x\} \equiv A $
    \item If $ M = \lambda x : B. N $ then  $ \Sigma; \Gamma, x:B \vdash C:\textbf{\textup{s}} $, $ \Sigma;\Gamma,x:B \vdash N:C $ and $ A \equiv \Pi x:B.C $
    \end{itemize}    
  \end{enumerate}
\end{theorem}

}

\end{document}